\documentclass[aps,prb,floatfix,amsmath,amssymb,preprint,eqsecnum,nofootinbib,superscriptaddress]{revtex4}
\usepackage{graphicx}
\usepackage{dcolumn}
\usepackage{bm}
\usepackage{epstopdf}
\usepackage{setspace}  
\newcommand{\beq}{\begin{equation}}
\newcommand{\eeq}{\end{equation}}
\newcommand{\ba}{\begin{array}{ccc}}
\newcommand{\ea}{\end{array}}
\newcommand{\nn}{\nonumber}
 \renewcommand{\d}{\partial}
\def\bea{\begin{eqnarray}}
\def\eea{\end{eqnarray}}

\def\<{\langle}
\def\>{\rangle}

\usepackage{amsmath}
\usepackage{amssymb}
\usepackage{color}
\begin{document}
\title{Interaction effects on 3D topological superconductors: surface topological order from vortex condensation, the 16 fold way and fermionic Kramers doublets.}

\author{Max A. Metlitski}
\affiliation{Kavli Institute for Theoretical Physics, UC Santa Barbara, CA 93106, USA. }

\author{Lukasz Fidkowski}
\affiliation{Department of Physics and Astronomy, Stony Brook University, Stony Brook, NY 11794-3800, USA.}

\author{Xie Chen}
\affiliation{Department of Physics, University of California, Berkeley, CA 94720, USA.}

\author{Ashvin Vishwanath}
\affiliation{Department of Physics, University of California, Berkeley, CA 94720, USA.}
\affiliation{Materials Science Division, Lawrence Berkeley National Laboratories, Berkeley, CA 94720, USA.}

\begin{abstract} Three dimensional topological superconductors with time reversal symmetry (class DIII) are indexed by an integer $\nu$, the number of surface Majorana cones,  according to the free fermion classification. The superfluid B phase of He$^3$ realizes the $\nu=1$ phase. Recently, it has been argued that this classification is reduced in the presence of interactions to $Z_{16}$.
This was argued from the symmetry respecting surface topological orders of these states, which provide a non-perturbative definition of the bulk topological phase. Here, we verify this conclusion by  focusing on the even index case, $\nu=2m$, where a vortex condensation approach can be used to explicitly derive the surface topological orders. We show a direct relation to the well known result on one dimensional topological superconductors (class BDI), where interactions reduce the free fermion classification from Z down to  Z$_8$.  Finally, we discuss in detail the fermionic analog of Kramers time reversal symmetry, which allows semions of some surface topological orders to transform as  $T^2=\pm i$.


\end{abstract}

\maketitle
\section{Introduction} 
Following the initial success in the prediction, experimental discovery and classification of free fermion topological phases\cite{FranzMolenkamp2013}, recent theoretical attention has turned to the effect of interactions. It has been now understood that the essential character of these states can be defined in a  wider context that includes strongly interacting systems (see Ref. \onlinecite{Senthil2014,TurnerVishwanath2013} for reviews).  In the process, three qualitatively new features were identified that are absent in the non-interacting problem:

 First, new topological phases appear that necessarily require the presence of interactions.  In fact, it is now understood that free fermion topological insulators and superconductors are examples of so-called``symmetry protected topological" (SPT) phases of matter. These phases have a bulk gap, and as their name suggests, are defined only in the presence of symmetry; once the symmetry is broken, a non-trivial SPT phase can be continuously connected to a local product state. This means that SPT phases have no ``intrinsic topological order," i.e.  they have only non-fractional excitations in the bulk. However, the non-triviality of an SPT phase is manifested by the presence of exotic edge states. Note that without interactions only SPT phases of fermions are possible, but in the presence of interactions SPT phases of bosons (or spins) also exist. While examples of such bosonic SPT phases were known in 1D \cite{Haldane,Chen1d}, a recent breakthrough has been the identification of SPT phases in 2D and 3D.\cite{ChenScience,Chen2011b, Chen2013, LevinGu,LuAV_SPT,AVTS,MetlitskibTI, Bi2013,Kitaev_pc, Kapustin2014,Kapustin2014a} 

Second, qualitatively new kinds of surface states may appear for SPT phases in 3D. In particular, the surface may be fully gapped while preserving all symmetries, if it realizes an appropriate form of intrinsic topological order. The symmetry action on the anyons of the topological order is of a form that is impossible to realize in a purely 2D system\cite{AVTS}. Examples of such surface states for SPT phases of bosons were discussed in Refs.~\onlinecite{AVTS, Burnell2013,Wang2013, MetlitskibTI, Bi2013, Cho2014, ChenBurnell2014}. Progress has also been made in identifying the corresponding  surface topological order (STO) for SPT phases of fermions, as well, including the ``conventional" non-interacting topological insulators and superconductors. Here, the symmetry respecting, gapped topologically ordered surface state is obtained by turning on strong interactions on the surface, while keeping the bulk of the system non- (or weakly) interacting. An SO(3)$_3$ non-Abelian surface topological order was proposed for the $\nu=1$ topological superconductor in Ref.~\onlinecite{FidkowskiChenAV}. Shortly after, the surface topological order for topological insulators was determined \cite{Metlitski2013,Wang2013a, Bonderson2013, Chen2014PRB, ChongScience, Senthil2014}, which was also found to be necessarily non-Abelian.   

Third, the free fermion classification of topological phases may be reduced in the presence of interactions. Two phases that appear to be topologically distinct at the level of free fermions, may in fact be  essentially the same phase on including interactions. That is, although on tuning from one phase to the other one always encounters a phase transition if the fermions are free, they can be adiabatically connected in the presence of interactions, and hence are the same phase. Indeed, it was shown that in 1D, topological superconductors (TSc's) in class BDI  (superconductors with time reversal symmetry, where $T^2 = +1$), which are  labelled by integers according to the free fermion classification, are reduced to just eight distinct phases, i.e. Z $\rightarrow$ Z$_8$  with interactions \cite{Fidkowski0, Fidkowski1,Turner1d,You2014a}. In the absence of interactions, the edge of a 1D TSc with label $\nu \in \mathrm{Z}$ supports $\nu$ stable Majorana zero modes. However, with interactions, the $\nu = 8$ edge can be fully gapped. Similar results hold for various 2D phases of fermions\cite{Qi,RyuZhang,LevinGuFerm, YaoRyu} including non-chiral topological superconductors with a global Z$_2$ symmetry.  In this last case, while a free fermion classification indicates that any number of counter propagating Majorana  edge modes with opposite Z$_2$ charge will be stable, adding interactions reduces this down to a Z$_8$ classification \cite{LuAV, LevinGuFerm}. In the above examples, the effect of interactions may be treated perturbatively to check for the stability of edge states. Moreover, the edge states are $0+1$ and $1+1$ dimensional systems and hence amenable to theoretical analysis even at strong coupling. One may ask a similar question regarding the stability of three dimensional topological phases. 

A particularly important example is provided by topological superconductors protected by time reversal symmetry (class DIII, where time reversal acts as $T^2=(-1)^F$, with $(-1)^F$ - the fermion parity). This is, indeed, the physical case of time reversal symmetric superconductors  with spin-orbit interactions. At the level of free fermion classification, there is an integer classification of topological superconductors in 3D, characterized by $\nu$ Majorana surface cones. Opposite signs of $\nu$ refer to Majorana cones with left or right chirality, which is well defined in the presence of time reversal symmetry. The B-phase of superfluid He$^3$, where the fermionic atoms pair to form an atomic superfluid, realizes the $\nu=1$ topological phase. A key question then is: are distinct values of $\nu$  truly different phases, or can they be smoothly connected in the presence of interactions? An elementary argument establishes that at least a Z$_2$ subgroup must survive interactions \cite{WangQi2011}, i.e. the phases with odd integer labels cannot be trivial. A perturbative analysis of Majorana cone surface states shows that interactions are irrelevant and hence the free fermion classification can survive weak interactions. However, this conclusion may be changed with strong interactions: perhaps, by turning on the interactions beyond a certain threshold, the surface can be driven into a trivial gapped, symmetry respecting phase. Answering this question seems formidable  since it appears to require a non-perturbative treatment of a 2+1D system. Initially, anomaly based arguments\cite{WangQi2011,Ryu2012,QiWitten2013} concluded that the entire Z classification was stable to interactions. However, more recently it was realized that the free fermion classification collapses to Z$_{16}$ \cite{Kitaev_pc,FidkowskiChenAV}. This was shown in Ref.~\onlinecite{FidkowskiChenAV} by using symmetry respecting surface topological order as a non-perturbative definition of the 3D topological phase, and identifying four distinct surface topological orders that lead to exactly $2^4$ phases and a Z$_{16}$ classification. Arguments for connecting these topological orders with specific $\nu$ were also given.

Here we will verify this conclusion by explicitly constructing the surface topological orders for {\em even} $\nu$ topological superconductors starting with the free fermion surface states and showing that they form a Z$_8$ group (combined with the odd $\nu$ this gives Z$_{16}$).  We note that during the completion of this work, other groups have used a similar vortex condensation approach in the context of topological superconductors with various symmetries \cite{Wang2014} and in different dimensions\cite{You2014}. The results derived here agree broadly with  these works and with Ref.~\onlinecite{FidkowskiChenAV}, although additional details of the surface topological order and its symmetry transformation properties are reported here.

The interacting generalizations of the free fermion classes AIII  (superconductors with time reversal and conservation of $z$-component of electron spin) and CI (superconductors with time reversal and full $SU(2)$ spin-rotation symmetry) are also reduced using similar arguments from Z to Z$_8$ and Z$_4$, as discussed in Ref.~\onlinecite{Wang2014}. An alternate way to view the reduction of topological phases is that interactions allow us to beat certain fermion doubling theorems that are derived assuming free fermions. Within a purely 2D free fermion model with time reversal symmetry, it is impossible to mimic the surface of the $\nu=16$ topological superconductor, i.e. to realize a state with 16 Majorana cones of the same `chirality'. However, this can be realized in an interacting system: one concrete recipe is to take a slab of the $\nu=16$ topological superconductor and gap one surface with interactions while preserving all symmetries. Now, reducing the system to two dimensions gives the desired state. Such a viewpoint was recently advocated in Ref.~\onlinecite{You2014} on extending to 4+1D topological phases, and interesting consequences for the  lattice regularization of the standard model of particle physics were derived.


Our construction of surface topological orders for even $\nu = 2m$ 3D topological superconductors will be based on the following strategy. We note that when $\nu$ is even, the system admits an enlarged `flavor' $U(1)$ symmetry. We first imagine driving the surface of the topological superconductor into a `superfluid' phase where this symmetry is spontaneously broken and the surface Majorana cones are gapped. This surface phase also breaks the time reversal symmetry, however, a combination of time reversal and a discrete rotation in the $U(1)$ group survives. Unlike the physical time reversal symmetry $T$, this remnant anti-unitary symmetry $S$ satisfies $S^2 = +1$.  We then imagine quantum disordering the surface superfluid by proliferating vortex defects. However, it turns out that for general $\nu$ the vortices are non-trivial: a vorticity $k$ defect traps $k m$ `chiral' Majorana zero modes in its core, and so resembles the edge of a 1D topological superconductor in class BDI. As we already noted,\cite{Fidkowski0, Fidkowski1} interactions can fully lift the degeneracy associated with the Majorana zero modes only when they come in multiples of $8$. Thus, for a $\nu = 16$ topological superconductor the elementary $k=1$ vortex is trivial. Its proliferation restores $U(1)$ symmetry, while preserving $S$, and thereby also restores $T$, giving rise to a topologically trivial symmetry-preserving gapped surface state. This proves that $\nu = 16$ 3D TSc is trivial. For smaller even $\nu$, the elementary vortex is non-trivial and cannot proliferate without breaking $S$, however, vortices with $k m \, \equiv \,0 \, (\mathrm{mod}\, 8)$ are trivial. Their proliferation gives rise to a symmetry respecting topologically ordered surface state, whose anyon content we explicitly determine. Our results are summarized in table \ref{tbl:answer}.

Some of these surface phases support semion excitations with unusual $T^2 = \pm i$ quantum numbers under time reversal symmetry. Usually, we are only able to define the action of time reversal symmetry locally and assign an anyon a precise value of $T^2$  when its statistics is unaffected by $T$. Here, although a semion is converted into an anti-semion under $T$, we are still able to define a generalized local time reversal symmetry, since the  semion and the anti-semion differ by a local electron excitation. This leads to a generalization of Kramers doublets,  where the two $T$-partners within the doublet have different fermion parity. We term such $T$-partners, `fermionic Kramers doublets.' We will discuss various examples of fermionic Kramers doublets in Section \ref{sec:pmi}. In addition to surface anyons, we will also recall more familiar contexts such as the edges of 1D topological superconductors and vortex defects of 2D topological superconductors, where a similar action of time reversal symmetry occurs.

This paper is organized as follows. We begin in Section \ref{3DTSc} with a brief introduction to topological superconductors in 3D with time reversal symmetry (class DIII) and point out that the phases labeled by even integers admit an enlarged U(1) symmetry that will prove useful. In Section \ref{8foldway} this symmetry is exploited to define vortices and map the problem to a well known prior result on the stability of 1D topological superconductors  (class BDI). In section \ref{sec:STO} we deduce the symmetry respecting topologically ordered surface phases via vortex proliferation.  Section \ref{sec:pmi} is devoted to studying fermionic Kramers doublets in more detail.

\section{Topological Superconductors in 3D} 
\label{3DTSc}
Topological superconductors protected by time reversal symmetry  with $T^2 = (-1)^F$ (class DIII) in 3D  have  $\nu$ gapless Majorana cones on their 2D surfaces. For example, the $\nu=1$ topological superconductor, realized by superfluid He$^3$ B, has gapless surface states described in the simplest case by the low energy dispersion:
\beq
{\mathcal H} _{\nu=1}=\chi^T \left (  p_x\sigma_z +p_y\sigma_x\right )\chi
\eeq 
where $\chi^T = \left ( \chi^\uparrow,\, \chi^\downarrow\right )$ is a two component Majorana field ($\chi^{\dagger} = \chi$) and $\vec{\sigma}$ are the Pauli matrices in the usual representation. Time reversal symmetry acts via:

\begin{eqnarray*}
T: \quad \chi^\uparrow &\rightarrow& \chi^\downarrow\\
\quad \chi^\downarrow &\rightarrow& - \chi^\uparrow
\end{eqnarray*} 
Clearly $T^2=-1$ when acting on the fermions. The mass term $\chi^T \sigma_y \chi$ is forbidden by time reversal symmetry. 

In the following we will consider topological superconductors where $\nu=2m$ is an even integer. The reason is that we will be able to introduce an artificial U(1) symmetry in those cases that will greatly aid the analysis. Eventually, we will break down the symmetry back to the physically relevant time reversal symmetry. 

First consider the case of $m=1$, when a {\em pair} of Majorana cones is present. Assuming that they have the same velocity and are centered in the same spot of the Brillouin zone, we have:
\begin{equation}
{\mathcal H} =\sum _{a=1}^2 \chi_a^T \left (  p_x\sigma_z +p_y\sigma_x\right )\chi_a
\label{H2}
\end{equation}
We can, therefore, enlarge the symmetry to include an O(2) (=U(1)) group, that involves rotations of the Majorana operators between the two flavors $a=1,\,2$. Combining these flavors into a complex fermion:
\begin{eqnarray}
\psi_\uparrow &=&\chi^\uparrow_1-i\chi^\uparrow_2\\ \nonumber
\psi_\downarrow &=&\chi^\downarrow_1-i\chi^\downarrow_2
\end{eqnarray}  
In these variables, the Hamiltonian (\ref{H2}) takes the form:
\beq
{\mathcal H} =\psi^\dagger\left (  p_x\sigma_z +p_y\sigma_x\right )\psi
\eeq
and is identical to the Hamiltonian of a topological {\em insulator} surface with a single Dirac cone and the chemical potential pinned to the Dirac point. However, there is an important distinction in the way in which  time reversal acts. Thus, while $\psi$ transforms under the $U(1)$ symmetry as,
\beq U(1):\quad \psi_\sigma \to e^{-i \alpha} \psi_{\sigma}, \label{eq:U1}\eeq
under time reversal it has the following unusual transformation where the U(1) charge is reversed:
\bea T: \quad \psi_\sigma \to \epsilon_{\sigma \sigma'} \psi^{\dagger}_{\sigma'} \nn\\
\quad \psi^{\dagger}_\sigma \to \epsilon_{\sigma \sigma'} \psi_{\sigma'} \label{eq:Tpsi}\eea
Note that $T^2 = (-1)^F$. However, $T$ commutes with $U(1)$ rotations, i.e. the total symmetry is $U(1)\times T$, unlike for the topological insulator where it is $U(1)\rtimes T$. In other words the U(1) `charge' here behaves like a component of spin as far as its time reversal properties are concerned. A corollary is that a vortex of this U(1) will remain a vortex under time reversal. In contrast, the vortex of the usual charge U(1) is converted into an anti-vortex under time reversal.
 
Now consider spontaneously breaking the flavor U(1) symmetry by inducing a `superfluid' on the surface, via the condensate of $O(x) = \epsilon_{\sigma \sigma'} \psi_{\sigma} \psi_{\sigma'}$.  At the mean-field (Bogolioubov-de Genne) level this condensation can be described by adding a term,
\beq \delta H = \Delta^{*} \, O(x) + \Delta \, O^{\dagger}(x) \label{eq:Delta}\eeq 
to the Hamiltonian. This will gap the Majorana cones. Clearly, since the Majorana cones  do not require the flavor U(1) for stability, this condensate must also break $T$-symmetry. Indeed, under $T$,
\beq T: \quad O(x) \to - O^{\dagger}(x)\eeq 
Intuitively this is clear: we mentioned that  in terms of the interplay with $T$, the flavor U(1) symmetry is like a `spin' rotatation, and so $O(x)$ is akin to a ferromagnetic order parameter, whose orientation is reversed by $T$.

 However, we can combine time reversal with a rotation by $\pi/2$ in the U(1) group. The resulting transformation $S$ is a symmetry of the system and may be viewed as a modified time reversal since it remains an anti-unitary operator:
\beq S = U_{\pi/2} T \label{eq:S}\eeq
Under $S$,
\beq S: \quad O(x) \to O^{\dagger}(x)\eeq
hence, the superfluid phase respects $S$. Note, however,  that in contrast to the original time reversal symmetry, we have  $S^2 = + 1$ and so the gapped fermionic Bogolioubov quasiparticle of the surface superfluid is a Kramers singlet. Below, we will study in detail the vortices of this surface superfluid.

Surface states with any even number of Majorana cones $\nu=2m$ can be treated in a similar manner, introducing an artificial $U(1)$ symmetry by grouping them in pairs. This results in a surface state with $m$ Dirac fermions $\psi_i$, ($i = 1 \ldots m$), where each $\psi_i$ transforms as (\ref{eq:U1}) under the common $U(1)$ symmetry and as (\ref{eq:Tpsi}) under $T$. The $m$ Dirac cones are stable at the free fermion level due to the presence of time reversal symmetry. One can then similarly spontaneously break the $U(1)$ symmetry by a condensate of $O(x) =  \sum_i  \epsilon_{\sigma \sigma'} \psi_{i \sigma} \psi_{i \sigma'}$. This condensate again simultaneously breaks $T$, but leaves the combination $S =U_{\pi/2} T$ preserved. 

\section{Vortices and the Eight Fold Way}
\label{8foldway}

Here, we introduce vortices into the surface superfluid of a topological superconductor with even index $\nu=2m$, and deduce their transformation properties under the modified time reversal symmetry $S$. 
\subsection{The $m=1$ surface with strength $k$ vortices.}
Let us begin with the $m=1$ surface. Imposing a static vortex configuration with vorticity $k$ in Eq.~(\ref{eq:Delta}), $\Delta(\vec{x}) = |\Delta (r)| e^{i k \theta}$, and solving the Bogolioubov-de Gennes equation, one finds $|k|$ Majorana zero modes $\gamma_\lambda$,  $\lambda = 1\ldots |k|$, localized in the vortex core.\cite{JackiwRossi,FuKane}  
Crucially, a vortex maps into a vortex rather than an anti-vortex under $S$, so one can study the transformation properties of the Majorana modes under $S$. As we show in appendix \ref{app:BdG}, the Majoranas transform in a chiral manner,
\beq S:\, \gamma_\lambda \to \mathrm{sgn}(k) \gamma_\lambda \label{eq:Sgamma}\eeq
Thus, a vortex resembles the end of a 1D topological superconductor in class BDI, with $S$ playing the role of a $T^2 = +1$ time reversal symmetry. As noted in the introduction, in the absence of interactions, such 1D TSc's are labeled by an integer $\nu \in \mathrm{Z}$, which counts the number of chiral Majorana zero modes at the edge. A vortex of strength $k$ is, thus, like an edge of a 1D TSc with $\nu = k$. The mass terms $i \gamma_{\lambda} \gamma_{\lambda'}$ are prohibited by $S$, so in the absence of interactions the Majorana modes are stable. However, as was shown in Refs.~\onlinecite{Fidkowski0, Fidkowski1} interactions can fully lift the degeneracy associated with the Majorana modes if they come in multiples of 8, breaking down the classification of 1D TSc's to Z$_8$. Let us review this result in the present context.

For an elementary $k = 1$ vortex, there is a single Majorana mode in the vortex core. As is well-understood, the stability of this zero mode is purely topological and does not rely on any symmetry. Indeed, the fermionic Hilbert space associated with a single Majorana mode is not well defined. However, if we have two vortices, each possessing a Majorana in its core then the two Majoranas $\gamma_1$, $\gamma_2$, form a two-dimensional Hilbert space, corresponding to the complex fermion mode $c = \frac{1}{\sqrt{2}}(\gamma_1 + i \gamma_2)$ being occupied or empty. If the vortices are far apart, these occupied and empty states are nearly degenerate as only a non-local fermion tunneling term $i \gamma_1 \gamma_2$ can split them. The presence of the Majorana mode endows the vortex with non-Abelian braiding statistics, as we will review shortly.

For a $k=2$ vortex, one finds a pair of Majorana modes in the vortex core. This pair of modes is not topologically protected, since a local mass term $i \gamma_1 \gamma_2$ lifts the degeneracy. However, the degeneracy is protected by $S$. Indeed, the mass term breaks $S$, furthermore, $i \gamma_1 \gamma_2$ is precisely the `local' fermion parity $(-1)^{F}$, counting whether the complex fermion mode $c = \frac{1}{\sqrt{2}}(\gamma_1 + i \gamma_2)$ in the vortex core is empty or filled. Now, $S (-1)^{F} S^{-1} = - (-1)^{F}$. Thus, each $k = 2$ vortex state must have a degenerate partner under $S$ with opposite fermion parity. We will refer to such defects as fermionic Kramers doublets, and will study them in more detail in section \ref{sec:pmi}. Representing $\gamma_1 = \sigma^x$, $\gamma_2 = \sigma^y$, $(-1)^F = -\sigma^z$, $S = \sigma^x K$, we see that for the $k  =2$ vortex, $S^2 = +1$. In contrast, for the $k = -2$ vortex, choosing an identical representation of $\gamma$'s, the opposite chirality of Majorana modes in Eq.~(\ref{eq:Sgamma}) gives,  $S = \sigma^y K$, so that $S^2 = -1$. Thus, for such fermionic Kramers doublets the value of $S^2$ is not tied to the degeneracy.  Note that the two states in each fermionic Kramers doublet differ by a Bogolioubov quasiparticle with $S^2 = +1$, so they have the same value of $S^2$.  One may ask if the value of $S^2$ is even physically meaningful in the present case. Indeed, a definite value of $S^2$ (or $T^2$) can only be assigned when the state and its time reversal partner differ by a local excitation. We are used to this excitation being a local boson, in which case one either has $S^2 = +1$ (ordinary Kramers singlet) or $S^2 = -1$ (ordinary Kramers doublet). However, an electron (Bogolioubov quasiparticle) is also local excitation, which leads to the possibility of a fermionic Kramers doublet with a well-defined value of $S^2$ (see section \ref{sec:pmi} for more details).

Next, consider the $k= 4$ vortex. The only perturbation in the ground state subspace allowed by $S$ is,
\beq \Delta H = U \gamma_1 \gamma_2 \gamma_3 \gamma_4 \eeq
$\Delta H$ is nothing but the fermion parity of the zero energy subspace. Therefore, the four-fold degenerate zero energy subspace splits into two doublets with opposite fermion parity. These doublets are actually (ordinary) Kramers doublets under $S$ and cannot be further split. (Note that in the present case $S$ preserves
the fermion parity).  Indeed, let us represent
$\gamma$'s by $4\times 4$ matrices: $\gamma_1 = \sigma^x$, $\gamma_2 = \sigma^y$, $\gamma_3 = \sigma^z \tau^x$, $\gamma^4 = \sigma^z \tau^y$. We see that $S$ can be implemented by $S = \sigma^x \tau^y K$ and $S^2 = -1$. The fermion parity $(-1)^F = \sigma^z \tau^z$ commutes with $S$. Note that even though each of the $k = 2$ vortices has $S^2 = +1$, when we fuse them together we obtain a $k = 4$ vortex with $S^2 = -1$. Likewise, fusing a $k = 2$ vortex and a $k= -2$ vortex, which have $S^2 = +1$ and $S^2 = -1$, respectively, gives a zero vorticity state with $S^2 = +1$. As we will show in section \ref{sec:pmi}, the fusion product of two fermionic Kramers doublets $1$, $2$, actually satisfies $S^2 = - S^2_1 S^2_2$, consistent with the present observations.

Finally, let's consider $k = 8$. Represent $\gamma_1 \ldots \gamma_4$ as above and $\gamma_5 = \sigma^z \tau^z \mu^x$, $\gamma^6 = \sigma^z \tau^z \mu^y$,
$\gamma^7 = \sigma^z \tau^z \mu^z \nu^x$, $\gamma^8 = \sigma^z \tau^z \mu^z \nu^y$. We can take $S = \sigma^x \tau^y \mu^x \nu^y K$. We see that $S^2 = 1$ and $S$ commutes with the fermion parity $(-1)^F = \sigma^z \tau^z \mu^z \nu^z$. Therefore, the degeneracy will generally be fully lifted and the $k=8$ vortex transforms trivially under $S$. 

Thus, we see that the transformation properties of vortices under $S$ remain invariant under a shift of vorticity $k \to k + 8$. So, we only need to discuss values $k = 0\ldots 7$. 
It  remains to discuss the $k = 3$ vortex, which carries three Majorana modes. Again, one of these Majoranas is protected topologically just as in the $k = 1$ case. However, all three modes are protected by $S$. The presence of these three Majorana modes is manifested in the following way. Imagine a pair of distant vortices with $k = 3$ and $k=-3$. The total of six resulting Majoranas will give rise to an 8-fold degenerate state. Out of this degeneracy, a 2-fold degeneracy is associated with opposite values of the overall fermion parity of the two-vortex state. The remaining four-fold degeneracy can be attributed to a (standard) local two-fold Kramers degeneracy of each vortex. In this sense, we may say that the $k=3$ vortex (and the $k = -3$ vortex) has $S^2 = -1$. (We make this notion more precise in section \ref{sec:pmi}.) On the other hand, if we have a pair of distant vortices with $k = 1$ and $k = -1$ then the system has only a two-fold degeneracy associated with the overall fermion parity, so the vortices carry no local Kramers degeneracy. We, thus, say that the $k = 1$ (and the $k = -1$) vortex has $S^2 = +1$. 

We list $S^2$ values deduced for all the vorticity sectors in table \ref{vortex_nu2} (bottom).

\subsection{Vortices on the surface of a $\nu=2m$ TSc.}
It is simple to generalize the above discussion to a 3D TSc with a general $\nu = 2m$. Considering a defect with vorticity $k$ on the superfluid surface, each flavor of Dirac cones supports $k$ Majorana modes in the vortex core, transforming as in Eq.~(\ref{eq:Sgamma}) under $S$. Thus, the vortex carries a total of $k m$ chiral Majorana modes, and transforms under $S$ as an end of a 1D TSc with $\nu = k m$. In particular, for $m = 2$ the elementary $k = 1$ vortex switches fermion parity under $S$, for $m = 4$ it is a (standard) Kramers doublet, and for $m = 8$ it is a trivial Kramers singlet. Therefore, for $m = 8$ the elementary vortex can proliferate, while preserving $S$. This proliferation restores the $U(1)$ symmetry and hence, simultaneously restores $T = U^{\dagger}_{\pi/2}  S$. One, thus, obtains a gapped, symmetry respecting surface phase with no intrinsic topological order, demonstrating the triviality of the $\nu = 16$ 3D TSc phase. This establishes the deep connection between 1D TSc's in class BDI and 3D TSc in class DIII, and the respective breakdown of the non-interacting classification, $\mathrm{Z} \to \mathrm{Z}_8$ (1D TSc's) and $\mathrm{Z} \to \mathrm{Z}_{16}$ (3D TSc's). For smaller values of $m$, the elementary vortex transforms non-trivially under $S$ and so cannot proliferate without breaking $S$, however, vorticity $k$ defects with $k m  \equiv 0 \, (\mathrm{mod}\,8\,)$ are trivial under $S$ and can proliferate, giving rise to a symmetry respecting, gapped surface with intrinsic topological order, which we will discuss in the next section.

\section{Surface Topological Order from Vortex Condensation}
\label{sec:STO}
 In this section, we will construct symmetry respecting topologically ordered surface states for 3D TSc's with $\nu = 2m$. We will obtain these by quantum disordering the surface superfluid via vortex proliferation, closely following the discussion of Metlitski {\em et al.}\cite{Metlitski2013} (see also Wang {\em et al.}\cite{Wang2013a}), who derived a STO for the fermionic topological insulators. We have described how vortices transform under the symmetry $S$ in the previous section; this, in particular, tells us which vortices can condense without breaking $S$. Another important property of the vortices that we will need to establish is their statistics - only vortices with bosonic statistics can proliferate. Vortex statistics can be inferred using the `slab trick' introduced in Ref.~\onlinecite{Metlitski2013} and reviewed below. We will find that the smallest $S$-trivial vortex (one with strength $k$, where $k m  \equiv 0 \, (\mathrm{mod}\,8\,)$) actually has bosonic statistics. We imagine condensing this vortex. This restores the $U(1)$ symmetry and preserves $S$, thereby also restoring $T = U^{\dagger}_{\pi/2}  S$. If $k > 1$, the resulting gapped surface state supports topological order. Finally, we will imagine a further surface phase transition, where the `artificial' $U(1)$ symmetry is broken (but $T$ is preserved), exposing the surface topological order relevant to the topological superconductors. We will follow this procedure in turn for each even $\nu$. The case $\nu  = 2$ is discussed in detail below; $\nu = 4$ and $\nu = 8$ are relegated to  appendix \ref{sec:48}; the remaining cases $\nu = 6, 10, 14$ are discussed in appendix \ref{app:2other}.

\begin{figure}[htbp]
\includegraphics[width=0.8\linewidth]{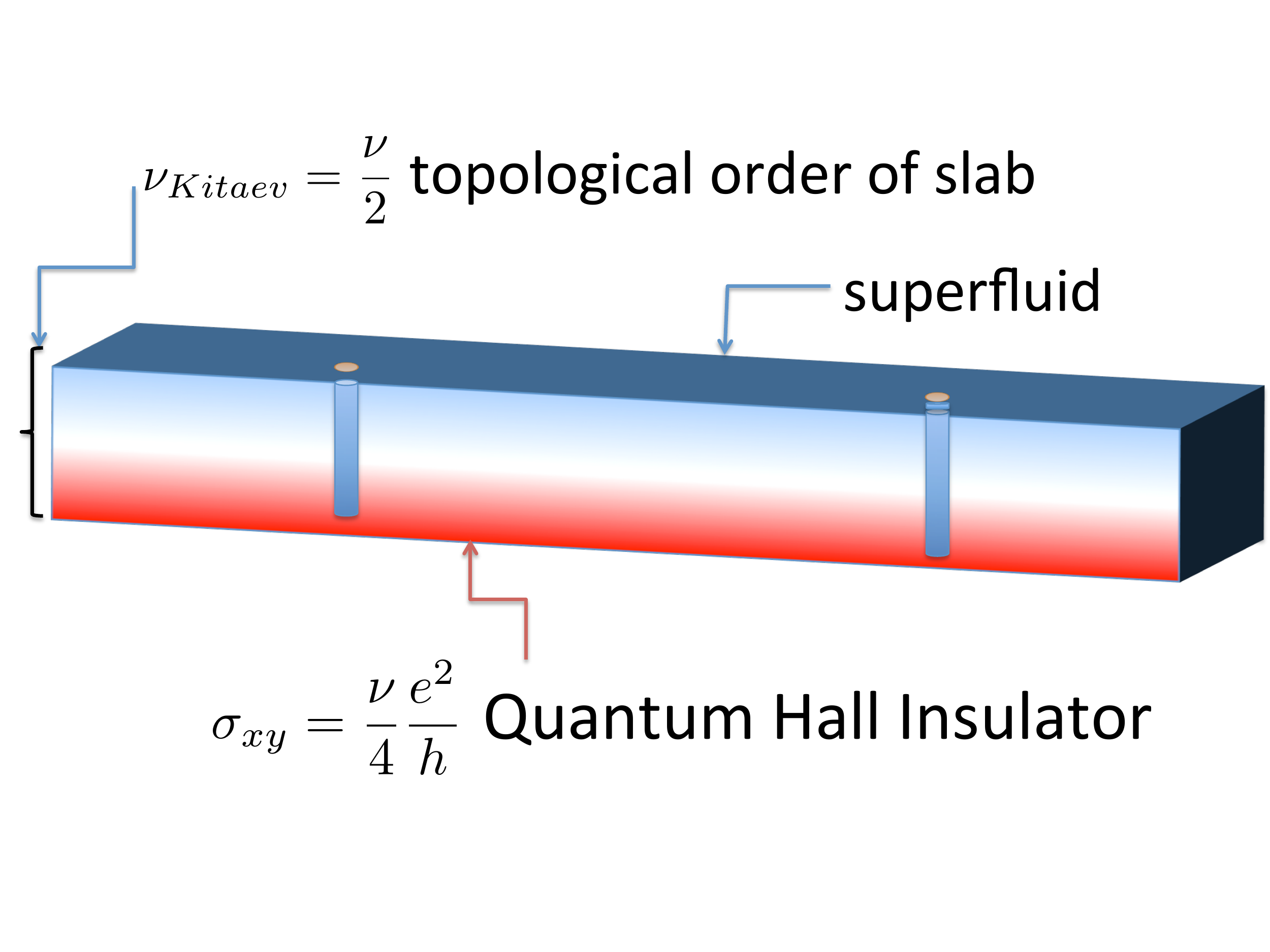}
\caption{{  Slab geometry with superfluidity on the top surface (which breaks both $U(1)$ and $T$ symmetry, but preserves the combination $S$ in Eq.~(\ref{eq:S})), and insulator on the bottom surface which breaks $T$, for a topological superconductor with $\nu=2m$.  Vortices on the top surface trap gauge flux that leaks down to the bottom surface. The statistics of well separated flux-vortex composites piercing the slab is described by the $\nu_{\rm {Kitaev}}=\nu/2$ topological order (Ising for $\nu_{\rm {Kitaev}} = 1$, $U(1)_4$ for $\nu_{\rm {Kitaev}} = 2$, $U(1)_2\times U(1)_2$ for  $\nu_{\rm {Kitaev}} = 4$). To extract the intrinsic, time reversal invariant vortex statisics associated with the top surface, the contribution of the bottom surface must be subtracted, resulting in vortex statistics described by a $\nu_{\rm{Kitaev}} \times U(1)_{-16/\nu}$ theory. Subsequently, the vortices of strength $16/\nu$ are condensed to give the surface topological order.}}
\label{RF}
\end{figure}


\subsection{The STO of the $\nu=2$ topological superconductor (single Dirac cone) from vortex condensation} 
\label{sec:nu2}

\subsubsection{Vortex Statistics}
\label{sec:vortstat}

Let us first tackle the case of the $\nu=2$ TSc's, that most closely parallels the discussion of the TI in Ref.~\onlinecite{Metlitski2013} (see also Ref.~\onlinecite{Wang2013}).  Let us begin by deducing the vortex statistics. Recall that to make the notion of vortex statistics (particularly, the Abelian part of the statistics) precise, we need to gauge the $U(1)$ symmetry by coupling the system to a weakly fluctuating gauge field $A_{\mu}$.  Vortices of strength $k$ now carry magnetic flux $khc/2e$. Next, consider a slab of the TSc with the opposite faces separated by a thickness much larger than the confinement length of the surface states. Thus, opposite surfaces are effectively decoupled. Imagine driving the top surface of the slab into a supefluid phase by a condensate of $O$, Eq. (\ref{eq:Delta}). In contrast, break $T$ on the bottom surface by a $U(1)$-preserving term,
\beq   \delta H = m \psi^{\dagger} \sigma^y \psi\eeq
This results in a fully gapped bottom surface whose response to the gauge field $A_{\mu}$ is characterized by a Hall conductivity $\sigma_{xy} = \frac12
 (e^2/h)$.
 Considering a vortex on the top surface, the associated magnetic flux will be confined to the vortex core near the top surface but will spread out in the bulk of the slab and on the bottom surface.

By solving the Bogolioubov-de Gennes equation one finds that the 1D interface on the TSc surface between the superfluid phase and the $\sigma_{xy} = 1/2$ phase supports a single chiral Majorana mode with central charge $c = 1/2$. The edge of our slab is precisely such an interface. Thus, viewing the entire slab as a 2D system, the presence of the chiral Majorana edge mode allows us to identify it with a $p+ip$ superconductor. The statistics of vortices in a $p+ip$ superconductor are known to be described by the Ising anyon theory. This theory has three topologically distinct sectors: ${I, \sigma, \psi}$.  Vortices with odd vorticity carry an odd number of Majorana zero modes and belong to the non-Abelian $\sigma$ sector, while vortices with even vorticity have an even number of Majorana zero modes and belong to sectors $I, \psi$, where $I$ is the trivial bosonic sector and $\psi$ - the Bogolioubov quasiparticle. The topological spins of the sectors are $(\theta_I,\,\theta_\sigma,\,\theta_\psi) = (0,\,\pi/8,\,\pi)$. 

It is clear that the topological spins of the vortices piercing the slab are sensitive to the way in which $T$ is broken on the bottom surface: if one breaks $T$ in the opposite way, the bottom surface has a Hall conductivity $\sigma_{xy} = -1/2$, the edge of the slab carries a Majorana mode moving in the opposite direction, and the slab is identified with a $p-ip$ superconductor. As a result, the odd strength vortices will have a topological spin $\theta_\sigma \to -\theta_\sigma = -\pi/8$. We are not actually interested in the statistics of the vortices associated with the entire slab. Rather, we would like to deduce the intrinisic contribution to the statistics coming from the top (supefluid) surface. To isolate these intrinsic statistics, we decompose the action of the slab as $S_{\mathrm{slab}} = S_{\mathrm{top}} + S_{\mathrm{bulk}} + S_{\mathrm{bottom}}$.  The slab as a whole is described by Ising anyon theory, $S_{\mathrm{slab}} = S_{\mathrm{Ising}}$. 
Note that during the vortex motion the bulk and the bottom surface are only affected via the magnetic flux emanating from the vortex, and so their contribution can be described by the effective actions $S_{\mathrm{bulk}}[A_{\mu}]$, $S_{\mathrm{bottom}}[A_{\mu}]$. The bulk action, $S_{\mathrm{bulk}}[A_{\mu}] \sim \int d^3x d\tau F^2_{\mu \nu}$ does not contribute to the vortex statistics. On the other hand, the Hall response of the bottom surface gives,
\beq S_{\mathrm{bottom}}[A_{\mu}] = \frac{i n}{4 \pi} \int d^2 x d \tau \epsilon_{\mu \nu \lambda} A_{\mu} \d_{\nu} A_{\lambda}, \quad n = 1/2 \label{eq:Hall}\eeq
Therefore, we deduce
\beq S_{\mathrm{top}} = S_{\mathrm{Ising}} - \frac{i}{8 \pi} \int d^2 x d \tau \epsilon_{\mu \nu \lambda} A_{\mu} \d_{\nu} A_{\lambda} \label{eq:Stop}\eeq
where the gauge field $A_{\mu}$ is constrained to follow the vortex motion via $j^v_{\mu} = \frac{1}{\pi} \epsilon_{\mu \nu \lambda} \d_{\nu} A_{\lambda}$, with $j^v_{\mu}$ - the vortex current. This constraint allows us to rewrite the second term in Eq.~(\ref{eq:Stop}) in terms of a dynamical Chern-Simons gauge field $a_{\mu}$ at level $k = -8$ coupled to the vortex current,
\beq S_{\mathrm{top}} = S_{\mathrm{Ising}} +\int d^2 x d \tau \left( -\frac{8 i}{4 \pi} \epsilon_{\mu \nu \lambda} a_{\mu} \d_{\nu} a_{\lambda} - i a_{\mu} j^v_{\mu}\right) \label{eq:Stop2}\eeq

Thus, the intrinsic vortex statistics is described by an $\mathrm{Ising} \times U(1)_{-8}$ anyon theory. This theory has topological sectors $X_{k}$, where $X\in \{1, \sigma, \psi\}$ denotes the charge in the Ising sector and 
the subscript $k$ denotes the charge in the $U(1)_{-8}$ sector, which coincides with the vorticity. Note that only a subset of all anyon types in the $\mathrm{Ising} \times U(1)_{-8}$ theory is realized by the vortices, since vortices of odd strength $k$ necessarily carry charge $\sigma$ in the Ising sector and vortices of even strength $k$ carry charges $1$ or $\psi$.  The resulting allowed topological sectors together with the corresponding topological spins, derived from $\theta_{X_k} = \theta_X -\pi \frac{k^2}{8}$, are listed in table \ref{vortex_nu2} (top). Following Ref.~\onlinecite{Chen2014PRB}, we call such a restriction of the $\mathrm{Ising}\times U(1)_{-8}$ anyon theory - the T-Pfaffian. Note that the topological properties are invariant under shifting the vorticity $k \to k + 8$. This coincides with the $k \to k+8$ periodicity of the vortex transformation properties under the time reversal-like symmetry $S$, discussed in section \ref{8foldway}. Let us check that the vortex statistics deduced above are consistent with the transformation properties under $S$.


\begin{table}[t]

\beq
\begin{array}{|c|c|c|c|c|c|c|c|c|}
\hline
k\rightarrow     & 0 & 1 & 2 & 3 & 4 & 5 & 6 & 7 \\ \hline
I    & +1 &   & -i &   & +1 &   & -i &   \\ \hline
\sigma  &   & \ 1 &   & -1&   & -1&   & \ 1 \\ \hline
\psi & -1&   & +i&   & -1&   & +i&  \\ \hline 

\end{array}\nn
\eeq
\beq
\begin{array}{|c|c|c|c|c|c|c|c|c|}
\hline
k\rightarrow     & 0 & 1 & 2 & 3 & 4 & 5 & 6 & 7 \\ \hline
S^2    & +1 & +1  & +1 & -1  & -1 & -1  & -1 & +1\\\hline
\end{array}\nn
\eeq

\caption{Vortex defects on the surface of a $\nu=2$ topological superconductor. Top: topological spins $e^{i \theta}$ of the vortices. The statistics are described by Ising$\times U(1)_{-8}$ theory. The column index is the flux $k\frac{hc}{2e}$, which coincides with the $U(1)_{-8}$ charge. The row index is the Ising charge.
Under $S$ time reversal $I_2\leftrightarrow \psi_2$ and $I_6\leftrightarrow\psi_6$. The other sectors are left invariant under $S$. Bottom: assignments of $S^2$ for the different vorticity sectors.}

\label{vortex_nu2}
\end{table}

Under $S$, the vorticity $k$ is left invariant (which is a significant point of difference from the case of the topological insulator surface). As under $S$ the topological spin, $e^{i \theta} \to e^{-i \theta}$, we must have, $S: I_2\leftrightarrow \psi_2$ and $I_6\leftrightarrow \psi_6$. This is consistent with our discussion in section \ref{8foldway}, where we found that the $k = 2$ and $k = -2 \sim 6$ vortices change their fermion parity under $S$. All the other topological sectors are mapped into themselves under $S$, again consistent with section \ref{8foldway}. 
Next, observe that $\psi_4$ is a fusion product  of $I_2$ and $\psi_2$,  which are exchanged by $S$ and have mutual semionic statistics. A very general argument\cite{Metlitski2013,Wang2013a,Chen2014PRB} then implies that $\psi_4$ must be a $S^2=-1$ fermion, again in agreement with section \ref{8foldway}.

Before proceeding to the vortex condensation to obtain the STO, we first note that the vortex statistics obtained here are identical to those derived by Metlitski {\em et al.} for the superconducting surface of a TI, except that time reversal acts very differently here, leaving fluxes invariant. Second, imagine a magnetic monopole of the $U(1)$ gauge field in the bulk of the $\nu = 2$ phase. From the Hall conductivity $\sigma_{xy} = 1/2$ of the $T$-broken surface state we conclude that the monopole must carry a half-odd-integer $U(1)$ charge, i.e. the bulk electromagnetic response is characterized by $\theta = \pi$, as in the TI.  We can study the passage of a magnetic monopole excitation from the vacuum into the bulk of the 3D system. This leaves behind a Dirac string flux of $hc/e$, which we identify with the $k=2$ vortex. The semionic/anti-semionic statistics of this vortex is intimately tied to the half-integer charge of the monopole as discussed in Ref.~\onlinecite{Wang2013a}.

\subsubsection{Vortex condensation: topological order}

Now, we consider condensing the strength 8 vortices $I_{8}$. These are trivial both topologically and in terms of their transformations under $S$. Thus, their proliferation restores the global $U(1)$ symmetry,  while preserving $S$. 
When the $k = 8$ vortices condense, the vortices $X_k$ with $k=0,\,1,\,2,\dots,\,7$, survive as $U(1)$-{\it neutral} anyon excitations, preserving their statistics. In addition to these neutral vortex descendants, the resulting surface phase also possesses charged boson excitations  $e^{im\phi}$, with fractional charge $m\frac{2e}{8}$, $m \in \mathrm{Z}$. (The `charge quantum' $q_{\mathrm{min}} = \frac{2e}{8}$ is dual to the flux $\Phi = 8 hc/2e$ of the condensing vortex; $q_{\mathrm{min}} \Phi/\hbar c  = 2 \pi$). The $m = 8$ excitation can be identified with a charge $2e$ Cooper pair $O(x)$, so the topologically distinct excitations have  $m=0,\,1,\,2,\dots,\,7$.  All anyons of the surface phase can be obtained by fusing a charged boson with a vortex descendant, and will be denoted as $X_{k,\,m}=X_k e^{im\phi}$. The charged boson $e^{i m \phi}$ and the vortex descendant $X_k$ have a `charge-flux' mutual statistics $e^{-2 \pi i m k/8}$. Thus, the topological spin of their fusion product is $\theta_{X_{k,\,m}} = \theta_{X_k} - 2\pi mk/8$. The resulting topological order has 96 distinct anyons, and is identical to $T_{96}$ in Ref. \onlinecite{Metlitski2013}. All anyons have trivial mutual statistics with a charge $e$ fermion, $\psi_{0,\,4}=\psi_0 e^{4 i\phi}$, which is identified with the electron $f$.

Before we proceed further, it is instructive to discuss how to drive a phase transition from the obtained topologically ordered surface phase back to the superfluid phase. This occurs via the condensation of the charge $e/4$ boson $e^{i \phi}$, which spontaneously breaks the $U(1)$ symmetry.  Due to the non-trivial mutual statistics $e^{-2 \pi i k/8}$ between $e^{i \phi}$ and $X_{k, m}$,  the condensation (logarithmically) confines all the anyons: $X_{k,m}$ go back to being strength $k$ vortex defects of the superfluid. To see this, note that in the presence of a $U(1)$ gauge field $A_{\mu}$, after $e^{i \phi}$ condensation, $X_{k, m}$ will bind magnetic flux $k hc/2e$ so that the statistical phase $e^{- 2\pi i k/8}$ acquired  by $e^{i \phi}$ upon going around $X_{k,m}$ is cancelled by the  electromagnetic Aharonov-Bohm phase. 

We now discuss the transformation properties of the STO under the restored time reversal symmetry $T = U_{-\pi/2} S$.  Let us first discuss how the various anyon sectors map into each other under $T$. Since all anyons carry a definite $U(1)$ charge, if $S$ sends anyon $a$ to anyon $\tilde{a}$, so does $T$. The vortex descendants $X_{k,0}$ preserve the transformation properties  they had in the superfluid phase: $T:\, I_2 \leftrightarrow \psi_2$, $I_6 \leftrightarrow \psi_6$, and all the other particles map into themselves. The transformations of the charged bosons $e^{i m \phi}$ are determined by the requirement that the $U(1)$ charge $q \to - q$ under $T$. Thus, $T: e^{i m \phi} \to e^{-i m \phi}$. 

Next, we come to the issue of $T^2$ assignments. For an anyon with charge $q$, the $T^2$ assignment can be obtained from the $S^2$ assignment, 
\beq T^2 = (U_{-\pi/2} S)^2= U_{-\pi} S^2 = e^{- i \pi q} S^2 \label{eq:TSrel} \eeq
where we've used the fact that $U$ and $S$ commute (we give a more careful proof of the result (\ref{eq:TSrel}) in section \ref{sec:pmi}). We, however, must remember that $T^2$ and $S^2$ assignments are only meaningful for anyons that transform locally, i.e. $a \to a$ or $a \to a \times f$, where $f$ is the electron. The first case, $a \to a$ is familiar: here $a$ can be a Kramers-singlet ($T^2 = +1)$ or a Kramers-doublet ($T^2 = -1$). The second case, $a \to a \times f$ is that of fermionic Kramers doublets, already mentioned in section \ref{8foldway} and discussed in more detail in section \ref{sec:pmi}. Thus, among the charged bosons $e^{i m \phi}$ the only anyon that can be assigned $T^2$ and $S^2$ is the charge $e$ particle, $e^{4 i \phi}$: the time reversal partners $e^{4 i \phi}$ and $e^{-4 i \phi}$ differ by a local Cooper pair $O(x) = e^{8 i \phi}$. Recalling that the transition to the $S$-preserving superfluid phase proceeds via  condensation of $e^{i \phi}$, we conclude that $e^{4 i \phi}$ must have $S^2 = +1$ and hence, by Eq.~(\ref{eq:TSrel}), $T^2 = -1$. More generally, an anyon $X_{k,m}$ preserves its $S^2$ value (if defined) after the transition into the superfluid phase, where the charge part $e^{i m \phi}$ `melts' and $X_{k,m}$ reduces to a vortex $X_k$. (From this point of view $e^{4 i \phi}$ melts into the zero vorticity sector $I_0$ after the transition and so, indeed, carries $S^2 = +1$). The $S^2$ assignments of superfluid vortices computed in section \ref{8foldway} are listed in table \ref{vortex_nu2} (bottom). Thus, from Eq.~(\ref{eq:TSrel}), the neutral vortex descendants $X_{k, 0} = X_k$ have $T^2 = S^2$, i.e. $I_0$, $\psi_0$ have $T^2 = +1$; $I_4$, $\psi_4$ have $T^2 = -1$;  $\sigma_{1,7}$ have $T^2 = +1$; $\sigma_{3,5}$ have $T^2 = -1$. Fusing the $T^2 = +1$ neutral fermion $\psi_0$ with the $T^2 = -1$ charge $e$ boson $e^{4 i \phi}$, we obtain the physical electron $f = \psi_0 e^{4 i \phi}$ with $T^2 = -1$, as required. 

Note that the neutral vortex descendants $I_2$ and $\psi_2$ ($I_6$ and $\psi_6$) can not be assigned a $T^2$ or $S^2$ value in the topologically ordered phase, despite having a well-defined $S^2 = +1$ ($S^2 = -1$) in the superfluid. Indeed, in the superfluid the $S$-partners $I_2$ and $\psi_2$ differ by the Bogolioubov quasiparticle $\psi_0$, which is just the local electron. Thus, in the superfluid they form a fermionic Kramers doublet. However, in the topologically ordered phase $\psi_0$ becomes a non-local anyon excitation, which is distinct from the electron $f = \psi_0 e^{4 i \phi}$, so $I_2$ and $\psi_2$ do not have a well-defined $T^2$ assignment. On the other hand, the charge $e/2$ semion $I_{2,2}$ and the charge $-e/2$ anti-semion $\psi_{2,-2}$ differ by the electron $f$ and do form a fermionic Kramers doublet, $T: \, I_{2,2} \leftrightarrow \psi_{2,-2}$. Upon transition to the supefluid phase, these reduce to vortices $I_2$ and $\psi_2$ and so inherit the assignment $S^2 = +1$. Therefore, from Eq.~(\ref{eq:TSrel}) we find $I_{2,2}$ has $T^2 = - i$ and $\psi_{2,-2}$ has $T^2 = +i$. Such unusual $T^2 = \pm i$ assignements were first introduced in Ref.~\onlinecite{FidkowskiChenAV} and are discussed further in section \ref{sec:pmi}. Note that $I_{2,2}$ and $\psi_{2,-2}$ have opposite values of $T^2$ consistent with them differing by a $T^2 = -1$ electron $f$. Also note that $I_{2,2}$ ($T^2 = -i$) and $\psi_{2,-2}$ ($T^2 = +i$) fuse to an (ordinary) Kramers-doublet $\psi_{4,0}$ ($T^2 = -1$). This is consistent with the general rule proved in section \ref{sec:pmi}: if anyon $a$ has fermionic Kramers parity $T^2_a$ and anyon $b$  has fermionic Kramers parity $T^2_b$ then the (ordinary) Kramers parity of the fusion product satisfies $T^2_{a \times b} = - T^2_a T^2_b$.  

Repeating the above argument, for $k = \pm 2$, $m = \pm 2$ we find a fermionic Kramers pair, $T:\, I_{k, m} \leftrightarrow \psi_{k, -m}$, where $I_{k, m}$ has $T^2 = (+i) (-1)^{(k+m)/4}$ and $\psi_{k, -m}$ has $T^2 = (-i) (-1)^{(k+m)/4}$.

In contrast to the case of the TI, it is not possible to further simplify this topological order without breaking $U(1)$ symmetry. We now do so to recover the STO of the TSc with only $T$ symmetry. To this end we condense the $T^2=+1$ charge $e$ boson, $I_{4,\,4}$. The resulting topological order, which we name $T_{24}$, has 24 particles,  
since this condensate confines all particles except those $X_{k,\,m}$ with $m={\rm even}$, and particles that differ by the condensate $I_{4,\,4}$ are considered equivalent. Thus, the resulting anyons are $T_{24} = \{X_{k,\, m}\}$ with $k=0,\,1,\,2,\dots,\,7$ and $m=0,\,2$. The electron $\psi_{0,4}$ becomes identified with $\psi_{4,\,0}$. (Note that this topological order is different from the 24 anyon STO obtained for TIs in Refs.~\onlinecite{Metlitski2013,Wang2013a}).





It is convenient to divide the anyons into two sets: $X_{k, 0}$ and $X_{k, 2}$. The first set $X_{k,\,0}=X_k$ originate from the neutral vortex descendants and form the T-Pfaffian topological order, with topological spins listed in table \ref{vortex_nu2} (top). The anyons in the second set can be written as $X_{k,\, 2} =  X_{k+2\,,0} \times s$, where $s = I_{-2,\,2}$ is a semion (topological spin $\theta = \pi/2$). Note that $s\times s = I_{-4,\,4} \sim I$, and $s$ has trivial mutual statistics with all anyons $X_k$.  Thus, we may write the STO as a direct product of two sectors,
\beq T_{24} = \{X_k\}\times \{1,\,s\} = {\rm T-Pfaffian} \times \{I,\,s\} \label{eq:T24}\eeq


\begin{table}[t]

\beq
\begin{array}{|c|c|c|c|c|c|c|c|c|}
\hline
k\rightarrow     & 0 & 1 & 2 & 3 & 4 & 5 & 6 & 7 \\ \hline
I    & +1 &   & \,\times\, &   & -1 &   &\, \times\, &   \\ \hline
\sigma  &   & \,\, \eta &   &-\eta  &   & -\eta&   & \,\, \eta \\ \hline
\psi & +1&   & \,\times\, &   & -1&   & \,\times\, &  \\ \hline 
\end{array}
\eeq

\caption{Transformation properties of the (T-Pfaffian)$_{\eta}$ topological order under $T$. 
Under $T$, $I_2\leftrightarrow \psi_2$ and $I_6\leftrightarrow\psi_6$. The other sectors are left invariant by $T$. The $T^2$ assignements are listed above (anyons for which $T^2$ is not defined are marked with a cross). The values $\eta = \pm 1$ correspond to two variants of the T-Pfaffian state. The anyon topological spins are identical to those in table \ref{vortex_nu2} (top).}

\label{tbl:TPfaffian}
\end{table}

\begin{table}[t]

\beq
\begin{array}{|c|c|c|c|c|}
\hline
     & I & s & f & s f  \\ \hline
e^{i \theta}   & +1 & +i  & -1 & -i  \\ \hline
T^2  &  +1 & +i \zeta  &  -1 &-i \zeta \\ \hline
\end{array}
\eeq

\caption{The semion-fermion theory, $SF_\zeta = \{I, s\} \times \{I, f\}$. Topological spins $e^{i \theta}$ and $T^2$ assignments. Under $T$, $s \leftrightarrow sf$, while $I$ and $f$ map into themselves. $\zeta = \pm 1$ correspond to two variants of the SF theory with distinct $T$-transformations.}

\label{tbl:SF}
\end{table}

We now discuss the action of time reversal symmetry on $T_{24}$. $T$ maps the T-Pfaffian sector onto itself, as summarized in table \ref{tbl:TPfaffian}. We use the notation of Ref.~\onlinecite{FidkowskiChenAV}, which introduced two variants of the T-Pfaffian topological order, denoted (T-Pfaffian)$_\eta$ with $\eta = +1$ and $\eta = -1$. These two variants differ by the Kramers parity of non-Abelian quasiparticles: for $\eta= +1$, $\sigma_{1,7}$ have $T^2 = +1$, while $\sigma_{3,5}$ have $T^2 = -1$; for $\eta = -1$, the values are reversed. With this notation, the T-Pfaffian sector of the above STO for the $\nu = 2$ TSc  is (T-Pfaffian)$_+$. As shown in appendix \ref{app:2other}, the $\nu  =10$ TSc has an identical $T_{24}$ surface topological order, Eq.~(\ref{eq:T24}), but with the (T-Pfaffian)$_-$ variant. 

Next, consider the action of $T$ on the $\{1, s\}$ sector. We find $T:\, s = I_{-2,2} \leftrightarrow \psi_{-2, -2} = s \times f$. Thus, $s$ and $s f$ form a fermionic Kramers doublet, with $s$ carrying $T^2 = +i$ and $s f$ carrying $T^2 = -i$. Note that in the decomposition (\ref{eq:T24}), the electron $f =\psi_{4,0}$ belongs to the T-Pfaffian sector. However, since the electron is local with respect to all anyons, we may equivalently write,
\beq T^{\nu = 2}_{24} = (\mathrm{T}-\mathrm{Pfaffian})_+ \times \mathrm{SF}_+\eeq
where the semion-fermion topological order SF is defined as SF$_\zeta = \{I,s\}\times \{I, f\}$. Thus, the action of time reversal on $T_{24}$ factors into its action on the T-Pfaffian and SF sectors. For future reference, we have introduced two variants of the semion-fermion topological order, labeled by the index $\zeta = \pm 1$. The $T^2$ assignement of the semion $s$ is $T^2 = i \zeta$, and of the anti-semion $s f$ is $T^2 = -i \zeta$. While the STO of the $\nu = 2$ TSc contains the SF$_+$ variant, as discussed in appendix \ref{app:2other}, the STO of the $\nu = -2 \sim 14$ TSc contains the SF$_-$ variant.

Next, we discuss how to further simplify the STO for the $\nu = 2$ TSc. However, for this purpose, it proves convenient to discuss the $\nu = 2, 6, 10, 14$ topological superconductors together.

\subsection{{  STO for other $\nu=\pm2 \,({\rm mod }\,8)$ topological superconductors; 16-fold way}}
The construction presented above can be easily followed to derive the STO for $\nu = 6$, $\nu = 10$ and $\nu = 14$ topological superconductors. The details are given in appendix \ref{app:2other} and the result is shown in table \ref{tbl:answer}.  We find that $\nu = 2$ and $\nu = 14 \sim -2$ respectively have (T-Pfaffian)${_+}\times\,$SF$_+$ and (T-Pfaffian)$_{+}\times\,$SF$_-$ surface topological order. On the other hand, $\nu = 6$ and $\nu = 10 \sim -6$ respectively have (T-Pfaffian)$_- \times \,$SF$_-$ and (T-Pfaffian)$_- \times \, $SF$_+$ surface topological order.


It has been argued in Ref.~\onlinecite{Chen2014PRB} that the two (T-Pfaffian)$_\eta$ states are connected via a surface phase transition either to a trivial state or to the STO of the 3D bosonic SPT phase with $T$ symmetry. The latter STO is the ordinary toric code $\{I, e, m, \epsilon\}$, with $e$, $m$ - $T^2 = -1$ bosons, and $\epsilon$ - a $T^2 = +1$ fermion. Following Ref.~\onlinecite{Wang2013}, we refer to a toric code with such unusual $T^2$ assignments as the eTmT state. Note that once physical $T^2 = -1$ electrons $f$ are present the eTmT state is equivalent to a 3 fermion state $\mathrm{FFF}= \{I, F_1, F_2, F_3\}$, where $F_1$, $F_2$, $F_3$ have the same fusion rules as $e$, $m$, $\epsilon$ in the toric code, but instead are all $T^2 = +1$ fermions. Indeed, $\mathrm{eTmT} \times \{I, f \} = \mathrm{FFF} \times \{I, f\}$ with the identification $e = F_1 f$, $m = F_2 f$, $\epsilon = F_3$. Thus, we will use the labels eTmT and FFF interchangeably. Note that two copies of eTmT can be connected via a surface phase transition to a trivial state.

Furthermore, the two (T-Pfaffian)$_\eta$ states were shown to differ by precisely the eTmT state.\cite{Chen2014PRB} Hence, one of them is connected via a surface phase transition to a trivial state and the other - to the eTmT state.  Thus, either the STO of $\nu = 2$ and $\nu = 14$  can be reduced to $\mathrm{SF}_+$ and $\mathrm{SF}_-$, respectively, while $\nu = 10$ and $\nu = 6$ can be reduced to $\mathrm{eTmT} \times \mathrm{SF}_+$ and $\mathrm{eTmT} \times \mathrm{SF}_-$, respectively; or vice-versa. Unfortunately, at the present time it is unclear for which value of $\eta$ the T-Pfaffian is connected to a trivial phase, so we cannot say which of the above two possibilities is realized. In appendix \ref{sec:nu8}, we will show that the $\nu = 8$ topological superconductor has precisely the eTmT surface topological order. This is consistent with STO of $\nu = 2$ and $\nu = 10$ ($\nu = 6$ and $\nu = 14$) phases differing by eTmT. 

Also observe that $\mathrm{SF}_+ \times \mathrm{SF}_-$ can be driven into a trivial phase, consistent with $\nu =2+   14 = 16$ ($\nu = 6 +10 = 16$) being trivial.\cite{FidkowskiChenAV} Indeed, write $\mathrm{SF}_+ \times \mathrm{SF}_- = \{I, s_+\} \times \{I, s_-\} \times \{I, f\}$. $s_+$ is a semion with  $T^2 = +i$, and $s_-$ is a semion with $T^2 = -i$.  Under $T: s_+ \leftrightarrow s_+ f$, $s_- \leftrightarrow s_- f$, so $s_+ s_- \to s_+ s_-$. By the rule for calculating the (ordinary) Kramers parity of the fusion product of two fermionic Kramers anyons, $T^2_{s_+ s_-} = - T^2_{s_+} T^2_{s-} = -1$. So $s_+ s_-$ is a $T^2 = -1$ fermion and $s_+ s_- f$ is a $T^2 = + 1$ boson. Condensation of this boson confines all the anyons and gives a trivial phase.

On the other hand, $\mathrm{SF}_+ \times \mathrm{SF}_+$ cannot be made trivial.\cite{FidkowskiChenAV} In section \ref{sec:nu4} of the appendix, we will show that $\mathrm{SF}_+ \times \mathrm{SF}_+$ precisely coincides with the surface topological order of the $\nu = 2+ 2 = 4$ phase, derived via vortex condensation.

By combining two $\nu = 4$ topological superconductors, we obtain a $\nu = 8$ TSc, whose topological order is a product of four SF$_+$ sectors: $\{I, s_1\} \times \{I, s_2 \} \times \{I, s_3\} \times \{I, s_4\} \times \{I, f\}$. Now $s_i s_j$ with $i \neq j$ is a $T^2 = +1$  fermion, so $s_1 s_2 s_3 s_4$ is a $T^2 = + 1$ boson. Condensing this boson reduces the STO to 
\beq T^{\nu = 8} = \{I, s_1 s_2, s_1 s_3, s_1 s_4 \} \times \{I, f\}\eeq
Letting $F_1 = s_1 s_2$, $F_2 = s_1 s_3$, $F_3 = s_1 s_4$, we see that $F_i$ are fermions with $T^2 = +1$ that realize the three-fermion toric code topological order FFF, which as we showed above is equivalent to the eTmT topological order in the presence of physical electrons. Thus,
\beq T^{\nu = 8}  = \mathrm{FFF} \times \{I, f\} = \mathrm{eTmT} \times \{I, f\}\eeq
In section \ref{sec:nu8} of the appendix, we will show that this is precisely the topological order of a $\nu = 8$ TSc deduced from vortex condensation. 

{  The deduced  STOs of all even $\nu$ topological superconductors are summarized in table \ref{tbl:answer}.}

Finally we note that combining two of the $\nu=8$ surface topological orders leads to a trivial surface state, implying that the $\nu=16$ bulk topological phase is actually the same as $\nu=0$ in the presence of interactions. This implies that interactions reduce the $Z$ free fermion classification down to $Z_{16}$.


\begin{table}[t]
\beq
\begin{array}{|c|c|c|c|c|}
\hline
     & I & e & m &  \psi  \\ \hline
e^{i \theta}   & +1 & +1 & +1 & -1  \\ \hline
T^2  &  +1 & -1  &  -1 &+1 \\ \hline
\end{array}
\eeq
\caption{The eTmT toric code topological order: topological spins $e^{i \theta}$ and $T^2$ assignments. All sectors map into themselves under $T$. This STO is realized by a 3D bosonic SPT phase with time reversal and by the $\nu = 8$ 3D fermion TSc.}
\label{tbl:eTmT}
\end{table}

\begin{table}[t]

\beq
\begin{array}{|c|c|}
\hline
 \nu    &  \mathrm{STO} \\ \hline
2   &    \mathrm{(T-Pfaffian)}_+\times\,\mathrm{SF}_+\\ \hline
4   &    \mathrm{SF}_+ \times \ \mathrm{SF}_+\\ \hline
6 & \mathrm{(T-Pfaffian)}_-\times\,\mathrm{SF}_-\\ \hline
8 & \mathrm{eTmT} \times \{I,f\}\\ \hline
10 & \mathrm{(T-Pfaffian)}_-\times\,\mathrm{SF}_+\\ \hline
12 & \mathrm{SF}_- \times \rm{SF}_-\\ \hline
14 & \mathrm{(T-Pfaffian)}_+\times\,\mathrm{SF}_-\\ \hline
16 & \mathrm{trivial}\\ \hline
\end{array}
\eeq

\caption{Surface topological order derived via vortex condensation for even $\nu$ topological superconductors. One of the (T-Pfaffian)$_{\pm}$ topological orders can be further reduced to a trivial state and the other - to the eTmT state, although at present it is not known which one is which. }

\label{tbl:answer}
\end{table}

\section{Fermionic Kramers doublets and $T^2=\pm i$ time reversal action} 
\label{sec:pmi}

In section \ref{sec:STO}, we have encountered an example of anyons $a$ that transform under time reversal  as $T: a \to a f$, where $f$ is the electron. Furthermore, we have claimed that such anyons can be assigned a definite value of $T^2$: one of the anyons $a$, $a f$ carries $T^2 = +i$ and the other $T^2 = -i$. As the two anyons differ by a $T^2 = -1$ electron $f$, their opposite values of $T^2$ appear consistent. We will call such anyon pairs $a$, $a f$ - {\it fermionic Kramers doublets}. Such unusual doublets were first discussed in Ref.~\onlinecite{FidkowskiChenAV}; here we further elaborate on this phenomenon. Our treatment closely follows Ref.~\onlinecite{Levin2012} where the notion of a local (ordinary) Kramers degeneracy was rigorously defined.

Let us first recall what it means for a many-body state to have a local Kramers degeneracy. Take a many-body state $|v\rangle$ with short-range correlations, i.e. one where $\langle v| O_1 O_2| v \rangle = \langle v| O_1| v \rangle \langle v| O_2| v \rangle$ for any two operators $O_1$, $O_2$ localized at widely separated points. Assume $|v \rangle$ has an even number of electrons. Imagine that under time reversal,
\beq T |v\rangle = b_1 b_2 |v \rangle \label{eq:Doublet}\eeq
where $b_1$ and $b_2$ are {\it bosonic} operators localized near distant points $1$ and $2$. These two points can be locations of anyon excitations or of classical defects (such as vortices or edges of a  1D system). Assume the normalization $||b_1 |v\rangle|| = ||b_2 |v \rangle|| = 1$. Define operators,
\beq T_1 = T b_2, \quad T_2 = T b_1 \label{eq:TaTb} \eeq
As was proved in Ref.~\onlinecite{Levin2012},
\beq T^2_1 |v \rangle = T^2_2 |v\rangle = \xi |v\rangle \label{eq:bosKramers}\eeq
with $\xi = \pm 1$. When $\xi = -1$, the defects at $1$ and $2$ are said to be {\it local} Kramers doublets. In this situation, the four states $|v\rangle$, $T_1 |v\rangle = - b_1 |v\rangle$, $T_2 |v\rangle = - b_2 |v\rangle$ and $T |v\rangle = b_1 b_2 |v\rangle$ are orthogonal and degenerate in energy. Furthermore, no local time reversal invariant perturbation can split this degeneracy. 

Now let us generalize the above local `bosonic' Kramers degeneracy to local fermionic Kramers degeneracy. Assume $|v\rangle$ again has short range correlations and satisfies,
\beq T |v\rangle = c_1 c_2 |v \rangle \label{eq:fDoublet}\eeq
but now with $c_1$, $c_2$ - {\it fermionic} operators localized near $1$ and $2$. Again, we assume $||c_1 |v\rangle|| = ||c_2 |v \rangle|| = 1$. (We further relax the assumption that $|v \rangle$ has an even number of electrons).  Defining 
\beq T_1 = T c_2, \quad  T_2 = T c_1, \label{eq:T1T2ferm}\eeq
we prove in appendix \ref{app:fKramers} that
\beq T^2_1 |v \rangle = \xi_1 |v \rangle, \quad T^2_2 |v \rangle = \xi_2  |v \rangle \label{eq:fKramers}\eeq
where $\xi_1 = \pm i$, $\xi_2 = \pm i$ and $\xi_1 \xi_2 = (-1)^{N_v+ 1}$, with $(-1)^{N_v}$ - the fermion parity of $|v\rangle$. Again, the states $|v\rangle$, $T_1 |v\rangle = \xi_2 c_1 |v\rangle$, $T_2 |v\rangle =- \xi_1 c_2 |v\rangle$, $T |v\rangle = c_1 c_2 |v\rangle$ are orthogonal and degenerate in energy. No local time reversal invariant perturbation can split this degeneracy. 

One can also compute the Kramers parity of defects at positions $1$ and $2$ for the other degenerate states. Not surprisingly, for $c_1 |v\rangle$ one finds $\xi'_1 = - \xi_1$, while $\xi'_2 = \xi_2$. On the other hand, $c_2 |v\rangle$ has $\xi'_1 = \xi_1$ and $\xi'_2 = -\xi_2$. Finally, for $c_1 c_2 |v\rangle$ both $\xi_1$ and $\xi_2$ switch sign. So, the defect and its local fermionic Kramers partner have opposite values of $T^2$, as expected.

Note that the relationship $\xi_1 \xi_2 = (-1)^{N_v+ 1}$ is somewhat peculiar: the Kramers parity of the entire state $|v\rangle$  is {\it minus} the product of Kramers parities of the two defects.  This fact holds also when computing the local {bosonic} Kramers parity of a defect obtained by fusing two fermionic Kramers defects. For instance, suppose we have fermionic Kramers defects at points $1,2,3,4$, so that
\beq T |v \rangle = c_1 c_2 c_3 c_4 |v\rangle\eeq
Grouping defects into pairs $(1,2)$, $(3,4)$, we see that each pair is either a bosonic Kramers singlet or a bosonic Kramers doublet. It is easy to show that
\beq T^2_{12} |v\rangle = -T^2_1 T^2_2 |v\rangle \label{eq:Tcomb}\eeq
where $T_{12} = T c_3 c_4$, $T_1 = T c_2 c_3 c_4$ and $T_2 = T c_1 c_3 c_4$. Indeed, assuming $|v\rangle$ has an even number of electrons, by Eq.~(\ref{eq:bosKramers}),
\bea T^2_{12} |v\rangle &=& T^2_{34} |v\rangle = (T c_1 c_2)^2 |v\rangle = T c_1 T T^{\dagger} c_2 T c_1 c_2 |v\rangle = - T c_1 T c_1 T^{\dagger} c_2 T c_2 |v\rangle \nn\\
&=& - (T c_1)^2 (-1)^F (T c_2)^2 |v\rangle = -  T^2_{234} T^2_{1 3 4} |v\rangle\eea
Now by Eq.~(\ref{eq:fKramers}) and discussion below it, $T^2_{234} |v\rangle = T^2_{1} |v\rangle$, $T^2_{1 3 4} = T^2_2 |v\rangle$ and we obtain Eq.~(\ref{eq:Tcomb}).

We would like to warn the reader that the value of fermionic Kramers parity is highly sensistive to the definion of $T$-operation. A replacement, $T \to T (-1)^F$, flips the sign of $T^2_1$ (i.e. $T^2_1 = i \leftrightarrow - i$). Thus, one needs to be careful to use the same definition of $T$ throughout.

Fermionic Kramers degeneracy can also occur in systems where $T^2 = +1$. (In fact, we've already encountered an example of this when discussing edge states of 1D topological superconductors in class BDI with $\nu = 2$).
In this case again $|v\rangle$ has fermionic Kramers degeneracy if it satisfies Eq.~(\ref{eq:fDoublet}). Again defining $T_1$ and $T_2$ as in Eq.~(\ref{eq:T1T2ferm}) one can show that 
\beq T^2_1 |v \rangle = \xi_1 |v \rangle, \quad T^2_2 |v \rangle = \xi_2 |v \rangle \label{eq:fermKp1} \eeq
but now with $\xi_1 = - \xi_2 = \pm 1$. The states $|v\rangle$, $T_1 |v\rangle = -\xi_1 c_1 |v\rangle$, $T_2 |v\rangle = \xi_2 c_2 |v\rangle$, $T|v\rangle$ are orthogonal and degenerate in energy. Note that unlike in the case of bosonic Kramers degeneracy, here states with local $T^2 = +1$ do not correspond to a singlet.   Furthermore, the defects at $1$ and $2$ have opposite $T^2$ although the overall state has $T^2 = +1$. Thus, the rule (\ref{eq:Tcomb}) for computing $T^2$ of the fusion product also holds here. Also all four degenerate states have the same values of $T^2_1$, $T^2_2$ consistent with the electron having $T^2 = +1$.

In this paper, we have enlarged the symmetry of topological superconductors from $T$ to $U(1)\times T$. Thus, the system posessed both a $T^2 = (-1)^F$ time reversal and a $S^2 = +1$ time reversal, related by
\beq T = U_{-\pi/2} S \eeq
where $U_\alpha$ is a rotation in the $U(1)$ group. In this case, the $T^2$ and $S^2$ values can be related via Eq.~(\ref{eq:TSrel}). Indeed, suppose we have a system with anyons $1$ and $2$ carrying electric charges $q_1$, $q_2$. Due to the $U(1)\times T$ symmetry, the charge $q \to - q$ under $T$. Now suppose we have,
\beq T |v\rangle = O_1 O_2 |v\rangle \eeq
Here, the operators $O_1$, $O_2$ can be bosonic or fermionic. Now, $O_{i}$ must lower the electric charge by $2 q_{i}$,
\beq U_\alpha O_i U^{\dagger}_\alpha = e^{- 2 i q_i \alpha} O_i \eeq
Computing the Kramers parity,
\bea T^2_1 |v\rangle  &=& (T O_2)^2 |v\rangle = U_{-\pi/2} S O_2 U_{-\pi/2} S O_2 |v\rangle = U_{-\pi/2} S e^{-\pi i q_2} U_{-\pi/2}  O_2 S O_2 |v\rangle \nn\\
&=& e^{\pi i q_2} U_{-\pi} (S O_2)^2 |v\rangle = e^{- \pi i q_1} S^2_1 |v\rangle  \eea
where in the last step we've used $U_{-\pi} |v\rangle = e^{-\pi i (q_1 + q_2)} |v\rangle$.


\subsection{Examples of $T^2 = \pm i$ defects}
We now give some simple explicit examples of defects with local fermionic Kramers degeneracy. One example is provided by the edge states of a 1D topological superconductor in class DIII (with $T^2 = (-1)^F$). There is a single non-trivial phase
in this class. A representative of the non-trivial phase is provided by combining two Kitaev chains: one for spin-up electrons and one for spin-down elecrons:
\beq H =  i t \sum_{i = 1,\sigma}^{N-1} \gamma^2_{i, \sigma} \gamma^{1}_{i+1, \sigma} \eeq
where the electron operators $c_{i \sigma}$ on each site are written as $c_{i \sigma} = \frac12 (\gamma^1_{i \sigma} + i \gamma^2_{i \sigma})$. With the usual transformation law for electrons, $T: c_\sigma \to \epsilon_{\sigma \sigma'} c_{\sigma'}$, the Majorana operators $\gamma^{1,2}_{i \sigma}$ transform as,
\beq T:\quad \gamma^{1}_{i \sigma} \to \epsilon_{\sigma \sigma'} \gamma^{1}_{i \sigma'}, \quad \gamma^{2}_{i \sigma} \to -\epsilon_{\sigma \sigma'} \gamma^{2}_{i \sigma'}\eeq
The ground state of the chain has $-i \gamma^2_{i \sigma} \gamma^{1}_{i+1, \sigma} = 1$, leaving two Majorana zero modes (one for each spin) at each end: $\gamma^1_{i = 1, \sigma}$ and $\gamma^2_{i = N, \sigma}$. Below, we drop the site index on these modes with the understanding that $\gamma^1_\sigma$ and $\gamma^2_\sigma$ refer to the left and right ends of the chain respectively. It appears that the right and left ends of the chain transform with opposite `chirality' under $T$, however, we can always redefine $\gamma^{2'}_\uparrow = \gamma^2_\uparrow$, $\gamma^{2'}_\downarrow = - \gamma^2_\downarrow$, so that the two ends transform identically under $T$. We use the primed variables below and drop the primes.
The fermion parity in the ground state subspace of the chain can be expressed as
\beq (-1)^F =  (-i \gamma^1_{\uparrow} \gamma^1_\downarrow) (-i \gamma^2_{\uparrow} \gamma^2_\downarrow) \label{eq:Fpar}\eeq

Focusing on one end of the chain, we see that the local fermion parity $(-1)^{F_1} = -i \gamma^{1}_\uparrow \gamma^1_\downarrow$ changes sign under $T$. Thus, the states with $(-1)^{F_1} = \pm 1$ are degenerate in energy: the Majorana modes cannot be lifted and the edge of the chain is a fermionic Kramers doublet. A quick way to see that the end has $T^2_1 = \pm i$ is to confine oneself to the Hilbert space of just the left edge and represent $\gamma^1_\uparrow = \sigma^x$, $\gamma^1_\downarrow = \sigma^y$, $(-1)^{F_1}  =\sigma^z$ and $T = \frac{1}{\sqrt{2}}(\sigma^x + \sigma^y) K$. Then $T^2 = - i \sigma^z = -i (-1)^{F_1}$. 

One can also use the more precise definition of local $T^2$ introduced in the previous section. Consider a state $|v\rangle$ of the entire chain that has short-range correlations. Then $|v\rangle$ will have well-defined values of left and right fermion parities $(-1)^{F_{1}} = -i \gamma^{1}_\uparrow \gamma^{1}_\downarrow$, $(-1)^{F_{2}} = -i \gamma^{2}_\uparrow \gamma^{2}_\downarrow$. As $T$ switches the fermion parity of each end, we can write
\beq T |v\rangle = e^{i \varphi} \gamma^1_{\uparrow} \gamma^2_{\uparrow} |v\rangle\eeq
where $e^{i \varphi}$ is some phase. Now, using the definition in the previous section, $T_1 = T \gamma^2_{\uparrow}$ and
\beq T^2_1 = T \gamma^2_{\uparrow} T \gamma^2_{\uparrow} = T \gamma^2_{\uparrow} T^{\dagger} (-1)^F \gamma^2_{\uparrow} = - \gamma^2_{\downarrow} \gamma^2_{\uparrow} (-1)^F =
i (-i \gamma^1_{\uparrow} \gamma^1_{\downarrow}) = i (-1)^{F_1}
\label{eq:DIII1D} \eeq
where we've used Eq.~(\ref{eq:Fpar}). (Note, there is a systematic sign difference between Eq.~(\ref{eq:DIII1D}) and the calculation using the `local' representation $T_1 = \frac{1}{\sqrt{2}}(\sigma^x + \sigma^y) K$. We can just absorb this minus  sign into the formal definition). 

Another example of $T^2 = \pm i$ defects is provided by vortices in a 2D topological superconductor in class DIII. Again, there is a single non-trivial phase in this class. A representative of this phase is obtained by putting spin up electrons into a $p+ip$ superconductor and spin down electrons into a $p-ip$ superconductor. A vortex in such a superconductor supports two Majorana zero modes $\gamma_{\uparrow}$, $\gamma_{\downarrow}$, with transformation properties $T: \, \gamma_\sigma \to \epsilon_{\sigma \sigma'} \gamma_{\sigma'}$. Thus, the vortex transforms in exactly the same way as an edge of a 1D topological superconductor in class DIII discussed above and has $T^2 = \pm i$. 

\subsection{Vortices on the superfluid surface of a $\nu = 2$ TSc}

In this section, we re-examine the transformation properties of vortices on the superfluid surface of a $\nu = 2$ TSc in class DIII, discussed in section \ref{8foldway}. As we noted, this surface phase has an anti-unitary symmetry $S$, with $S^2 = +1$. A vortex with vorticity $k$ carries $|k|$ Majorana zero modes $\gamma_\lambda$, which transform according to Eq.~(\ref{eq:Sgamma}) under $S$. As we already pointed out, vortices with $k = \pm 2$ have fermionic Kramers degeneracy. Indeed, consider nucleating a $k = 2$ and a $k = -2$ vortex out of the vacuum. Denote the Majorana modes on the $k = 2$ vortex as $\gamma_{1,2}$ and the Majorana modes on the $k = -2$ vortex as $\bar{\gamma}_{1,2}$. The local fermion parities $-i \gamma_{1} \gamma_{2}$ and $-i \bar{\gamma}_1 \bar{\gamma}_2$ switch under $S$. Thus, the action of $S$ on the two-vortex state takes the form,
\beq S |v\rangle = \gamma_1 \bar{\gamma}_1 |v\rangle \eeq
Defining $S_{k = 2} = S \bar{\gamma}_1$ and $S_{k=-2} = S \gamma_1$, we find $S^2_{k = 2} = - 1$ and $S^2_{k = -2} =  +1$, consistent with our discussion in section \ref{8foldway}.  Applying the rule (\ref{eq:Tcomb}), we learn that the $k = 4$ vortex has $S^2 = -1$, i.e. is a bosonic Kramers doublet, again consistent with section \ref{8foldway}. 

Finally, we can clarify in which sense the $k= 1$ vortex is a bosonic Kramers singlet and the $k = 3$ vortex is a bosonic Kramers doublet. Let us nucleate a $k = 1$, $k = -1$ vortex pair out of the vacuum. Denote the corresponding Majorana modes as $\gamma$, $\bar{\gamma}$. Strictly speaking, the resulting state with definite fermion parity $-i \gamma \bar{\gamma} = \pm 1$ has non-local correlations as, $\langle \gamma \bar{\gamma}\rangle \neq \langle \gamma\rangle \langle \bar{\gamma} \rangle$. However, it has local correlation functions for all bosonic observables, which is sufficient to define bosonic Kramers degeneracy. Since $- i \gamma \bar{\gamma}$ is invariant under $S$, we find,
\beq S |v\rangle = e^{i \varphi} |v\rangle\eeq
and the $k  =1$, $k =-1$ vortices are bosonic Kramers singlets. Next, imagine nucleating a $k  =3$, $k = -3$ vortex pair out of the vacuum with corresponding Majorana modes $\gamma_\lambda$, $\bar{\gamma}_\lambda$, $\lambda = 1, 2, 3$. The degenerate subspace can be labeled by quantum numbers $-i \gamma_1 \gamma_2$, $-i \bar{\gamma}_1 \bar{\gamma}_2$, $-i \gamma_3 \bar{\gamma}_3$.  Any state where these quantum numbers are fixed has short-range correlations of all bosonic observables. Now, under $S$, $-i \gamma_1 \gamma_2$ and $-i \bar{\gamma}_1 \bar{\gamma}_2$ switch sign, while $- i \gamma_3 \bar{\gamma}_3$ remains invariant. Thus, the action of $S$ on any such state is,
\beq S |v\rangle = e^{i \varphi} (\gamma_1 \gamma_3) (\bar{\gamma}_1 \bar{\gamma}_3) |v\rangle\eeq
Note that $(\gamma_1 \gamma_3)$  and $(\bar{\gamma}_1 \bar{\gamma}_3)$  are bosonic operators localized on the two vortices. Defining $S_{k = 3} = S \bar{\gamma}_1 \bar{\gamma}_3$, we find $S^2_{k = 3} = -1$. It is in this sense that the $k = 3$, $k = -3$ vortices are bosonic Kramers doublets. 


{
 \subsection{Monopoles in the bulk of a $\nu = 2$ TSc with $U(1)$ symmetry}
Yet another example of fermionic Kramers defects is provided by magnetic monopoles in the bulk of a $\nu = 2$ TSc with an enlarged $U(1) \times T$ symmetry. As we already discussed in section \ref{sec:vortstat}, the response of this phase to a $U(1)$ gauge field $A_{\mu}$ is characterized by a $\theta$ term with $\theta = \pi$. Thus, magnetic monopoles of $A_{\mu}$ with flux $h c /e$ carry a charge $q = n + \frac12$, $n \in Z$. Due to the $U(1) \times T$ symmetry, the magnetic flux stays invariant under $T$, while the charge $q \to - q$. Thus, a charge $\frac12$ monopole is converted into a charge $-\frac12$ monopole under $T$, i.e. the time reversal partners differ by a charge $1$ electron and form a fermionic Kramers doublet. (Note that for this notion to be precise we must treat the magnetic flux as a background non-dynamical field. Once the gauge field is allowed to fluctuate, the electron becomes a non-local excitation and the value of $T^2$ becomes ill-defined.) The precise value of $T^2$ parity of the two states within the doublet can be deduced as follows. Drive the surface of the $\nu = 2$ TSc into the superfluid phase and imagine dragging a magnetic monopole across the surface. As noted in section \ref{sec:vortstat}, this process nucleates a strength $k = 2$ (flux $hc/e$) vortex on the surface. This vortex is a fermionic Kramers doublet under the time reversal-like symmetry $S$ and has $S^2 = + 1$. Thus, according to the discussion below Eq.~(\ref{eq:fermKp1}), the monopole in the bulk must carry $S^2 = -1$. Therefore, from Eq.~(\ref{eq:TSrel}), the monopole with charge $q$ carries $T^2 = - e^{-\pi i q}$. Similarly, the monopole with opposite flux has $T^2 = + e^{- \pi i q}$. }


\section{Conclusions}
In this paper we have systematically derived the surface topological order for the even index 3D topological superconductors, using the vortex condensation technique. The results confirm the essential picture described in Ref.~\onlinecite{FidkowskiChenAV} (see also Refs.~\onlinecite{ChongScience,Wang2014,Bi2014,You2014}), as well as the collapse of the integer classification down to Z$_{16}$, but allows for a more detailed specification of the STO corresponding to a particular topological superconductor. 

We have also discussed in detail the concept of fermionic Kramers doublets, some aspects of which have been touched upon in recent work \cite{QiHughes2009,Fidkowski0,Turner1d}. These can exist in systems built out of fermions (electrons or Bogoliubov quasiparticles), where certain excitations appear as doublets whose components differ in their fermion parity. Time reversal symmetry switches members of the doublets, which implies that it is a fermionic operator. We pointed out several examples of this phenomenon and the unusual physics associated with it. When the underlying fermions are Kramers singlets (as in class BDI), the fermionic Kramers doublet excitations have $T^2=\pm1$, and in both cases a two fold degeneracy is present. An example is furnished by the edge of a one dimensional topological phase in class BDI with topological index $\nu=2$. Furthermore, combining two $T^2=+1$ fermionic Kramers doublets leads, via the anti-commutation relation for fermions, to a bosonic Kramers doublet with $T^2=-1$! On the other hand when the underlying fermions are Kramers doublets (as in class DIII), the fermionic Kramers doublet excitations have $T^2=\pm i$, and again in both cases a two fold degeneracy is present. The surface topological order of the $\nu=2$ TSc in class DIII has semions and their time reversed partners that transform according to this pattern. We note that this provides an unusual example where we can define the local time reversal action for a particle that is neither  a boson nor a fermion. 

 In closing we would like to draw attention to some open questions. The vortex condensation approach provides a powerful route to connecting the free fermion surface states and surface topological orders, but is only applicable to the even $\nu$ topological phases. What about the case of odd $\nu$? A surface topological order, SO(3)$_3$, which is non-Abelian, but supports just  a single non-trivial anyon (aside from the electron) has been proposed as a root state for the odd $\nu$ in Ref. \onlinecite{FidkowskiChenAV}. While this has been shown to be a topological superconductor with the right thermal Hall conductivity at surface domain walls, a direct connection to a Majorana fermion surface state remains open.   Also, we observe that the T-Pfaffian state, which was  discussed in the context  of the topological insulator STO \cite{Chen2014PRB,Bonderson2013}, makes an appearance in the present work, as well. It was noted in Ref. \onlinecite{Chen2014PRB} that this STO appears in two flavors ${\rm T-Pfaffian}_\pm$, which differ by the STO of a 3D bosonic SPT phase. One of the flavors is the STO of the 3D fermionic topological insulator, which, in the absence of charge conservation, is a trivial state (i.e. can be realized in 2D).  However, which of the two this is remains to be established, and will lead to a simplification in table \ref{tbl:answer} as well.  
\section{Acknowledgements}
AV would like to acknowledge the Perimeter Institute where part of this work was performed, and support from NSF - DMR 1206728. XC is supported by the Miller Institute for Basic Research in Science at UC Berkeley. MM is supported in part by the National Science Foundation under Grant No. NSF PHY11-25915.

\appendix
\section{Bogolioubov de-Gennes equation for vortices}
\label{app:BdG}
In this appendix we show that a strength $k$ vortex on the superfluid surface of a $\nu  =2$ TSc carries $k$ chiral Majorana modes.
The (continuum) Bogolioubov-de Gennes Hamiltonian on the surface reads,
\beq H_{BdG}= \psi^{\dagger} (-i \d_x \sigma^z - i \d_y \sigma^x) \psi + \Delta^*(x) \psi_{\uparrow} \psi_{\downarrow} + \Delta(x) \psi^{\dagger}_{\downarrow} \psi^{\dagger}_{\uparrow} \eeq
Going to the Nambu notation,
\beq \Psi  = (\psi_{\uparrow}, \psi_{\downarrow}, \psi^{\dagger}_{\downarrow}, -\psi^{\dagger}_{\uparrow}), \quad \Psi^{\dagger}  = (\psi^{\dagger}_{\uparrow}, \psi^{\dagger}_{\downarrow}, \psi_{\downarrow}, -\psi_{\uparrow})\eeq
\beq H_{BdG} = \frac12 \Psi^{\dagger} h_{BdG} \Psi \eeq
where
\beq h_{BdG} = (-i \d_x \sigma^z - i \d_y \sigma^x) \tau^z + \Delta^*(x) \frac{\tau^x + i \tau^y}{2} + \Delta(x) \frac{\tau^x - i \tau^y}{2}\eeq
with $\tau$ matrices operating in the Nambu space.
The Nambu operators satisfy, $\Psi^{\dagger} = \tau^y \sigma^y \Psi$. Defining the anti-unitary charge-conjugation operator (in the single particle space)
\beq C = \tau^y \sigma^y K\eeq
with $K$ - complex conjugation, we have, $\{C, h_{BdG}\} = 0$ and $C^2 = 1$. 

Under time reversal, 
\beq T: \quad \Psi \to i \tau^y \Psi\eeq
and under $S$,
\beq S: \quad \Psi \to i \tau^x \Psi\eeq
Thus, on the single-particle level, the anti-unitary symmetry $S$ is represented as 
\beq S = i \tau^x K\eeq
and $[S, h_{BdG}] = [S, C] = 0$. Note that $S C = - \tau^z \sigma^y$ is a unitary symmetry, with $(SC)^2 = 1$ and $\{S C, h_{BdG}\} = 0$. 

Let $\phi_\lambda$ be eigenstates of $h_{BdG}$ with eigenvalue $\lambda$. To each positive energy level $\phi_{\lambda}$, $\lambda > 0$, there corresponds a negative energy level $\phi_{-\lambda} = C \phi_\lambda$. 
Moreover, $C$ maps the zero energy subspace into itself. Since $C$ is anti-unitary and $C^2 = 1$, without loss of generality we can choose the zero energy eigenstates to satisfy $C \phi_\lambda = \phi_\lambda$. Thus, $\Psi$ can be expanded as
\beq \Psi(x) = \sum_{\lambda > 0} \left( c_\lambda \phi_\lambda(x) + c^{\dagger}_{\lambda} (C \phi_\lambda)(x)\right) + \frac{1}{\sqrt{2}}\sum_{\lambda = 0} \gamma_{\lambda} \phi_\lambda(x) \eeq
with $\gamma^{\dagger}_\lambda = \gamma_\lambda$ and $\{\gamma_{\lambda}, \gamma_{\lambda'}\} = 2 \delta_{\lambda \lambda'}$.  

Since $S C$ maps the zero energy subspace into itself, the zero energy eigenstates can be choosen to be eigenstates of $S C$ (with eigenvalues $\eta_\lambda =\pm 1$). Moreover, since $[SC , C] = 0$, this can be done while preserving the property $C \phi_\lambda = \phi_\lambda$.   
Hence, $\phi_\lambda$ also carries an eigenvalue $\eta_\lambda$ under $S$. Let $N_+$ be the number of zero energy states with $\eta_\lambda = +1$ and $N_-$ - with $\eta_\lambda = -1$. An index theorem\cite{Weinberg} states that
\beq N_+ - N_- = k\eeq
with $k$ -the winding number of the vortex. When the vortex is rotationally symmetric, all zero energy eigenstates have the same $\eta_\lambda = \rm{sgn}(k)$.\cite{JackiwRossi}

We also point out that since $[S, h_{BdG}] = 0$,  the finite energy eigenstates $\phi_\lambda$ can be chosen to satisfy $S \phi_\lambda = \phi_\lambda$.  

With the above convention, we conclude that the fermion operators transform under $S$ as,
\bea S \gamma_\lambda S^{-1} = \eta_\lambda \gamma_\lambda\ \nn\\
S c_\lambda S^{-1} = c_\lambda \eea
Note that $c_\lambda$ can be rewritten in terms of Majorana operators as $c_\lambda = \frac12 (\gamma^+_\lambda + i \gamma^-_\lambda)$. Then,
 \beq S \gamma^+_\lambda S^{-1} = \gamma^+_\lambda, \quad S \gamma^-_\lambda S^{-1} = - \gamma^-_\lambda\eeq
Thus, if we cut off our Hilbert space at some finite energy, we have a total of $\tilde{N}_+$ Majoranas with $S \gamma_\lambda S^{-1} = \gamma_{\lambda}$
and $\tilde{N}_-$ Majoranas with $S \gamma_\lambda S^{-1} = - \gamma_{\lambda}$, and $\tilde{N}_+ - \tilde{N}_- = k$. The transformation properties of the vortex under $S$
depend only on this difference, and so can be computed by focusing just on the zero modes.

\begin{table}
\beq
\begin{array}{|c|c|c|c|c|c|c|c|c|}
\hline
\nu \downarrow, \,\, k\rightarrow     & 0 & 1 & 2 & 3 & 4 & 5 & 6 & 7 \\ \hline
2    & +1 & +1  & +1 & -1  & -1 & -1  & -1 & +1\\\hline
6    & +1 &  -1 & -1 & +1 & - 1 & + 1 & + 1 & -1\\ \hline
10  & + 1 & -1 & +1 & +1 & -1 & + 1 & -1 & -1\\ \hline 
14  & + 1 & + 1 & -1 & - 1 & -1 & -1 & + 1 & + 1\\ \hline
\end{array}\nn
\eeq

\caption{$\nu = 2, 6, 10, 14$. Assignments of $S^2$ for the different vorticity sectors $k$. }

\label{tbl:oddm}
\end{table}

\section{$\nu = 2, 6, 10, 14$}
\label{app:2other}
Here we discuss the STO for topological superconductors with $\nu = 6$, $\nu = 10$, $\nu  = 14$ ($\nu = 2m$, $m - \rm{odd}$). The procedure is the same as for the $\nu = 2$ case. The first step is to deduce the vortex statistics on the superfluid surface using the slab trick. It is easy to see that vortex statistics are identical to the $\nu =2$ case. Indeed, the field content of $\nu = 6, 10, 14$ differs from $\nu = 2$ by a multiple of 2 Dirac cones. In the slab construction, condense the Cooper pairs $O(x) = \sum_{i = 1}^{m} \epsilon_{\sigma \sigma'} \psi_{i \sigma} \psi_{i \sigma'}$ on the top surface. On the bottom surface,  break the $T$-symmetry as,
\beq \delta H = m \psi^{\dagger}_1 \sigma^y \psi_1 + m \sum_{i = 2, 4, \ldots}^{m-1} (\psi^{\dagger}_{i} \sigma^y \psi_i - \psi^{\dagger}_{i+1} \sigma^y \psi_{i +1})\eeq
i.e. use opposite signs of mass term for the extra pairs of Dirac cones. As a result, these pairs do not contribute to the Hall conductivity $\sigma_{xy}$ of the bottom surface. They also give rise to pairs of counter-propagating Majorana edge modes that can be gapped out. Thus, they don't change the index  $\nu_{\rm{Kitaev}}$ of the slab. Therefore, the vortex statistics are the same as in the $\nu =2$ case. However, the transformation properties of vortices under $S$ are different. A strength $k$ vortex now transforms as the edge of a 1D BDI superconductor in class $\nu = k m$. The resulting $S^2$ values are summarized in table \ref{tbl:oddm}. Following the same procedure as in the $\nu = 2$ case, we obtain a (T-Pfaffian)$_\eta \times$SF$_\zeta$ surface topological order. The $S^2$ value of the $k = 1,7$ vortices translates into the $T^2$ value of the neutral vortex descendants $\sigma_{1, 7}$ and determines the $\eta$-parity of the T-Pfaffian sector. Thus, $\nu = 2, 14$ give rise to (T-Pfaffian)$_+$ and $\nu = 6, 10$ give rise to (T-Pfaffian)$_-$. Likewise, the $S^2$ value of the $k = 6$ vortex is inherited by the semion $s = I_{6} e^{2 i \phi}$ and translates into $T^2_{s} = (-i) S^2_{s}$. So, we find that $\nu = 2, 10$ give rise to $\rm{SF}_+$ and $\nu = 6, 14$ give rise to $\rm{SF}_-$.  


\section{Surface Topological Orders for $\nu=4$ and $\nu=8$}
\label{sec:48}
\subsection{The STO of the $\nu=4$ topological superconductor (double Dirac cone) from vortex condensation} 
\label{sec:nu4}
This case has twice the field content of the topological insulator, $\psi_{i\sigma}$, $i = 1,2$. Let us begin by using the slab trick to deduce the vortex statistics of the surface supefluid. As in section \ref{sec:nu2}, we drive the top surface of the slab into a superfluid phase by condensing,  $O = \sum_{i = 1}^2 \epsilon_{\sigma \sigma'} \psi_{i \sigma} \psi_{i \sigma'}$, and drive the bottom layer into a $T$-breaking insulator via
\beq  \delta H = m \sum_{i = 1}^2 \psi^{\dagger}_i \sigma^y \psi_i\eeq
The bottom surface has a Hall conductivity $\sigma_{xy} = 1 \times (e^2/h)$. The edge of the slab now carries two chiral Majorana modes, so the slab as a 2D system is identified with two copies of a $p+ip$ superconductor, $\nu_{\rm{ Kitaev}} = 2$. The statistics of vortices in a $\nu_{\rm{ Kitaev}} = 2$ superconductor are described by a $U(1)_4$ theory with anyon content $I_{n}$, $n = 0, 1, 2, 3$ and topological spins $\theta_{n} = \pi n^2/4$. The fermion $I_2$ is the Bogolioubov quasiparticle. Even strenth vortices belong to $n = 0, 2$ sectors and odd strength vortices -  to $n  =1,3$ sectors.
As in section \ref{sec:nu2}, to extract the intrinsic vortex statistics in the superfluid phase, we have to subtract out the contribution of the bottom surface. The Hall conductivity $\sigma_{xy} = 1$ of the bottom surface now translates into an additional factor of $U(1)_{-4}$ for the intrinsic vortex statistics, so overall the vortices can be described by a $U(1)_4 \times U(1)_{-4}$ theory with anyon content $I_{n, k}$, where $n = 0, 1, 2, 3$ is the charge in the $U(1)_4$ sector, and $k$ is the charge in the $U(1)_{-4}$ sector, which coincides with the vorticity. Note that $(n,k)$ must satisfy $n+k={\rm even}$. This simply reflects the fact that vortices with even strength $k$ belong to $n = 0, 2$ sectors and vortices with odd strength $k$ belong to $n = 1, 3$ sectors. The self statistics (topological spin) of a vortex $I_{n,\,k}$ is $\theta_{n,\,k}=\pi \left (\frac{n^2}4 - \frac{k^2}4\right )$. The resulting anyon content, with associated topological spins is displayed in Table \ref{table_nu4_vortex}.

An alternative way to deduce the vortex statistics is to again use the slab trick, but now breaking $T$ on the bottom surface via,
\beq  \delta H = m (\psi^{\dagger}_1 \sigma^y \psi_1 - \psi^{\dagger}_2 \sigma^y \psi_2) \eeq
i.e. use opposite signs of the mass $m$ for the two Dirac cones. The bottom surface now has $\sigma_{xy} = 0$. The edge of the slab carries two counter-propagating Majorana modes, which can generally be gapped out. Thus, the slab viewed as a 2D system is a non-chiral superconductor, $\nu_{Kitaev} = 0$. The vortices in this case realize the simple toric code $\{I, e, m, \psi \}$, where $e$, $m$ are bosons and $\psi$ is a fermion, which corresponds to the Bogolioubov quasiparticle. The odd strength vortices belong to $e$ and $m$ sectors and the even strengths vortices belong to $I$, $\psi$ sectors. Since the bottom surface now has $\sigma_{xy} = 0$, it does not contribute to the vortex statistics. Thus, the intrinsic vortex statistics is described just by the toric code. To keep track of the vorticity, we will still denote the anyons as $X_{k}$, where $X \in \{I, e, m, \psi\}$ and $k$ is the vorticity. Identifying $I_{0,0} = I_0$, $I_{2,0} = \psi_0$, $I_{1,1} = e_1$, $I_{3,1} = m_1$, $I_{0,2} = \psi_2$, $I_{2,2} = I_2$, $I_{1,3} = m_3$ and $I_{3,3} = e_3$, we see that the vortex fusion rules and topological spins are the same as in the $U(1)_4\times U(1)_{-4}$ description. Thus, in the $\nu =4$ case there is nothing exotic about the vortex statistics: they are the same as in a 2D $s$-wave superconductor and periodic under shifting $k \to k + 2$. However, as we show below, the transformations of vortices under the time reversal symmetry $S$ are unusual and only periodic under shifting $k \to k + 4$.

\begin{table}[t]

\beq
\begin{array}{|c|c|c|c|c|}
\hline
n\downarrow\, k\rightarrow     & 0 & 1 & 2 & 3  \\ \hline
0    & +1 &   & -1 &      \\ \hline
1  &  & +1 &  & +1    \\ \hline
2 & -1&   & +1 &  \\ \hline 
3 & & +1&& +1\\ \hline
\end{array}
\quad\quad
\begin{array}{|c|c|c|c|c|}
\hline
n\downarrow\, k\rightarrow     & 0 & 1 & 2 & 3  \\ \hline
0    & I &   & \psi &      \\ \hline
1  &  & e &  & m    \\ \hline
2 & \psi&   & I &  \\ \hline 
3 & & m& & e\\ \hline
\end{array}
\nn
\eeq
\beq
\begin{array}{|c|c|c|c|c|}
\hline
\, k\rightarrow     & 0 & 1 & 2 & 3  \\ \hline
S^2  & +1 & +1 & -1  & -1       \\ \hline
\end{array}
\nn
\eeq

\caption{Vortex defects $I_{n,k}$ on the surface of a $\nu=4$ topological superconductor. Top (left): topological spins $e^{i \theta}$ of the vortices. The statistics are described by a $U(1)_4 \times U(1)_{-4}$ theory. The column index is the flux $k\frac{hc}{2e}$, which coincides with $U(1)_{-4}$ charge. The row index is the $U(1)_4$ charge. Under $S$ time reversal, $I_{1,1} \leftrightarrow I_{3,1}$ and $I_{1,3} \leftrightarrow I_{3,3}$; the other vortices map into themselves under $S$. Top (right): An equivalent representation of vortices in terms of anyons of toric code. Bottom:  $S^2$ assignments of different vorticity sectors. }
\label{table_nu4_vortex}
\end{table}
The transformations of vortices $I_{n,\,k}$ under $S$ time reversal symmetry follow from section \ref{8foldway}. First, the vorticity $k \to k$ under $S$. A vortex with strength $k$ carries $2 k$ chiral Majorana zero modes and transforms like the edge of a $\nu = 2 k$ 1D TSc in class BDI. The Bogolioubov quasiparticle $I_{0,2}$ is a Kramers singlet $S^2 = +1$. Vortices with $k  = 1$ belong to the $\nu = 2$ BDI class and so transform as a fermionic Kramers doublet $S:\, I_{1,1}\leftrightarrow I_{3,1}$ with $S^2 = +1$ (note that $I_{1,1}$ and $I_{3,1}$ differ precisely by the Bogolioubov quasiparticle $I_{2,0}$). Similarly, vortices with $k = 3$ belong to the $\nu = 6$ BDI class and transform like a fermionic Kramers doublet $S:\, I_{1,3} \leftrightarrow I_{3,3}$ with $S^2 = -1$. Vortices $I_{0,2}$ and $I_{2,2}$ with $k = 2$ belong to the $\nu =4$ BDI class and so are (ordinary) Kramers-doublets with $S^2 = -1$. Note, that the $S$-transformations are consistent with the vortex statistics. In particular,  $I_{1,\,1}$and  $I_{3,\,1}$ have mutual semionic statistics and are exchanged by $S$, so by the result of Refs.~\onlinecite{Metlitski2013,Wang2013a,Chen2014PRB} their fusion product $I_{0,2}$ must be a Kramers doublet, as we, indeed, find. The $S$ transformations are invariant under shifting the vorticity $k \to k+4$. In particular, $I_{0,4}$ is a (trivial) $S^2 = +1$ boson.

Before proceeding, we note that unlike for the $\nu = 2$ case, the magnetic monopole in the bulk of the $\nu = 4$ phase carries an integer charge, as can be seen from the integer Hall conductivity of the $T$-broken surface insulator. Flux $hc/e$ ($k = 2$) vortex excitations are generated on the surface when a magnetic monopole passes from the outside to the inside of the system. Both vortices with $k=2$ transform as $S^2=-1$, hence we conclude that a neutral monopole has $\,T^2 = -1$. That is the neutral monopole possesses a Kramers degeneracy.   The $T^2 = -1$ assignment only makes sense here since the magnetic monopole  transforms into itself (and not an anti-monopole) under time reversal symmetry.  Again this is a consequence of the $U(1)\times T$ symmetry (as opposed to the $U(1)\rtimes T$ symmetry of topological insulators, where no such possibility is allowed). Furthermore, the monopole must be a boson. Indeed, when a neutral monopole passes through the surface, the vortex it nucleates must have trivial mutual statistics with all other vortex defects: i.e. it must be the boson $I_{2,2}$ ($I_2$ in toric code notation). Therefore, the monopole must also be a boson (actually, it was shown in Ref.~\onlinecite{ChongScience} that the external monopole is a boson in {\it any} fermion insulator with no topological order).  

Next, we consider condensing the $k = 4$ vortices $I_{0,4}$, which are both topologically trivial and $S$-trivial. This restores the $U(1)$ symmetry and leads to fractional charge boson excitations $e^{im\phi}$ with charge $m\frac{2e}{4}$, $m \in \mathrm{Z}$. $e^{4 i \phi}$ is a charge $2 e$ boson that we identify with a Cooper pair, so only $m = 0, \,1,\, 2,\, 3$ are topologically distinct. The quasiparticles now are $I_{n,\,k}e^{im\phi}$, $n = 0\ldots 3$, $k = 0\ldots 3$, $m = 0\ldots 3$, $n + k = \rm{even}$, and the statistics now includes a term that describes braiding of flux and charge, $\theta_{n,\,k,\,m} = \theta_{n,\,k} -2\pi mk/4$. \ 
The electron is identified as the charge $e$ fermion $f = I_{2,\,0}e^{2 i\phi}$, which braids trivially with all the other anyons.  
Thus, the resulting topological order $T^{\nu = 4}_{32}$ has 32 particles. Let us re-write the topological order in a more transparent form. Observe that the anyons $Y_{k, m} \equiv I_{k, k} e^{i m \phi}$, $k = 0\ldots 3$, $m  = 0 \ldots 3$, form the topological order of a $\mathrm{Z}_4$ gauge theory. Indeed, $I_{k,k}$ and $e^{i m \phi}$ are bosons with mutual statistics $-2 \pi m k/4$, so can be thought of as $Z_4$ fluxes and $Z_4$ charges, respectively. Any anyon on the surface can be written either as $Y_{k,m}$ or $Y_{k,m} \times f$. Thus,
\beq T^{\nu = 4}_{32} = \{Y_{k, m}\} \times \{I, f\} = \mathrm{Z}_4 \times \{I, f\}\eeq
This conclusion is not surprising: as we saw, the vortices of the surface superfluid have the same statistics as in a 2D s-wave superconductor, where condensation of a strength $k$ vortex leads to a $\mathrm{Z}_{k}$ topological order. 

The transition back to the superfluid phase occurs via condensation of the boson $e^{i \phi}$, upon which anyons $I_{n, k} e^{i m \phi} $ reduce to superfluid vortices $I_{n, k}$.

Next, we discuss the transformation properties of anyons under time reversal. Since time reversal sends charge $q \to -q$, we have $T: e^{i m \phi} \to e^{-i m \phi}$. The transformation properties of the vortex descendants $I_{n, k}$ are inherited from the superfluid phase: $T:\, I_{1,1} \leftrightarrow I_{3,1}$, $T: \, I_{1,3} \leftrightarrow I_{3,3}$ and all other $I_{n,k}$ map into themselves. The $T^2$ assignments of anyons can be worked out from the vortex $S^2$ assignments in table \ref{table_nu4_vortex} using Eq.~(\ref{eq:TSrel}). Thus, the neutral fermion $I_{2,0}$ is an (ordinary) Kramers singlet ($T^2 = +1$), $I_{0,2}$ and $I_{2,2}$ are (ordinary) Kramers doublets ($T^2 = -1$). The charge $e$ boson $e^{2i \phi}$ reduces to the zero vorticity sector upon the transition into the superfluid, so carries $S^2 = +1$ and $T^2 = -1$, i.e. it is an (ordinary) Kramers doublet under $T$. Fusing $e^{2 i \phi}$ with the neutral fermion $I_{2,0}$ we get the  $T^2 = -1$ electron $f$, as required.  The charge $e/2$ anti-semion $I_{1,1} e^{i \phi}$ and the charge $-e/2$ semion $I_{3,1} e^{-i \phi}$ are exchanged by $T$ and differ by the electron $f$: hence, they form a fermionic Kramers doublet. Both reduce to the $k = 1$ vorticity sector upon the transition back to the superfluid, and so inherit the assignment $S^2 = +1$, which translates into $T^2 = - i$ for $I_{1,1} e^{i \phi}$ and $T^2 = + i$ for $I_{3,1} e^{-i \phi}$. More generally, we find fermionic Kramers pairs $T:\, I_{n,\, k} e^{i m \phi} \leftrightarrow I_{n + 2, k} e^{-i m \phi}$ where $n = 1,\, 3$; $k = 1, \, 3$; $m = 1, \, 3$ with $T^2 = (+i) (-1)^{(k + m)/2}$ for $I_{n,\,k} e^{i m \phi}$. 

Now, we consider breaking the $U(1)$ symmetry, to recover the generic STO corresponding to the TSc. In order to do this we condense the charge $e$ particle $Y_{2,2} = I_{2, 2}e^{2 i\phi}$, which is a $T^2=+1$ boson, and hence its condensation preserves $T$.  The particles $I_{n, k} e^{i m \phi}$ that survive this condensation necessarily have $n+m=\rm even$. The resulting phase has 8 distinct anyons: $I_{n,\, k}$, $n = 0,\, 2$; $k = 0,\, 2$ and $I_{n,\, k} e^{i \phi}$, $n = 1,\,3$; $k = 1,\,3$. This anyon content can be conveniently written as,
\beq T^{\nu = 4}_8 = \{I, s_1\} \times \{I, s_2\} \times \{I, f\} \label{eq:T8}\eeq
where $s_1 = I_{1,3} e^{i \phi}$, $s_2 = I_{3,3} e^{i \phi}$ are two $\theta = \pi/2$ semions, and $f = I_{2,0} e^{2 i \phi} \to I_{0,2}$ is the physical electron. Note that $s_1 \times s_1 = s_2 \times s_2 = I_{2,2} e^{2 i \phi} \to I$; and $s_1$, $s_2$ have trivial mutual statistics, so the resulting surface topological order is, indeed, a direct product (\ref{eq:T8}). 

The transformation properties under $T$ are inherited from the $32$ anyon theory. We find fermionic Kramers doublets $T:  s_1 \leftrightarrow s_1 f$, $\,T: s_2 \leftrightarrow s_2 f$, where the semions $s_1$, $s_2$ have $T^2 = +i$, while the anti-semions $s_1 f$, $s_2 f$ have $T^2 = -i$. In addition, the fermion $s_1 s_2$ has $T^2_{s_1 s_2} = - T^2_{s_1} T^2_{s_2} = +1$ and the boson $s_1 s_2 f$ has $T^2 = -1$. Thus, the resulting STO $T_8$ is precisely two copies of the semion-fermion theory SF$_+$ in table \ref{tbl:SF}:
\beq T^{\nu = 4}_8 = \rm{SF}_+ \times \rm{SF}_+ \label{eq:T8exp}\eeq
This is fully consistent with the $\mathrm{(T-Pfaffian)}_+ \times \mathrm{SF}_+$ STO derived for the $\nu = 2$ TSc in section \ref{sec:STO}, since as shown in Ref.~\onlinecite{Chen2014PRB}, two copies of  $\mathrm{T-Pfaffian}_\eta$ topological order can be driven to a trivial phase via a surface phase transition.

Not surprisingly, repeating the above arguments for the $\nu = 12 \sim -4$ TSc gives a STO consisting of two copies of $\mathrm{SF}_-$ theory.





\subsection{The STO of the $\nu = 8$ topological superconductor (four Dirac cones) from vortex condensation}
\label{sec:nu8}
Finally, let us turn to the case of the $\nu=8$ topological {superconductor}. This case has 4 Dirac cones, $\psi_{i \sigma}$, $i = 1 \ldots 4$; thus, four times the field content of the topological insulator. Let us begin by deducing the vortex statistics using the slab trick. Condense the Cooper pairs $O = \sum_{i = 1}^4\epsilon_{\sigma \sigma'} \psi_{i \sigma} \psi_{i \sigma'}$ in the top layer of the slab, and drive the bottom layer into a $T$-breaking insulator via
\beq  \delta H = m \sum_{i = 1}^4 \psi^{\dagger}_i \sigma^y \psi_i\eeq
The bottom surface now has a Hall conductivity $\sigma_{xy} = 2 (e^2/h)$. The edge of the slab carries 4 chiral Majorana modes, so the slab as a 2D system is identified with 4 copies of a $p+ip$ superconductor, $\nu_{\rm Kitaev} = 4$. Vortex statistics in a $\nu_{\rm Kitaev} = 4$ superconductor are governed by a $U(1)_2 \times U(1)_{2}$ theory, with anyons $I_{n_1, n_2}$, $n_1 = 0,1$; $n_2 = 0,1$, where $n_1$ is the charge under the first $U(1)_2$ sector and $n_2$ is the charge under the second $U(1)_2$ sector. The topological spin of $I_{n_1,n_2}$ is $\theta_{n_1, n_2} = \frac{\pi}{2} (n^2_1 + n^2_2)$. The Bogolioubov quasiparticle is the fermion $I_{1,1}$. Vortices of strength $k$ belong to sectors $I_{0,0}$, $I_{1,1}$ for $k$ - even, and $I_{1,0}$, $I_{0,1}$ for $k$ - odd. Subtracting out the contribution of the $\sigma_{xy} = 2$ bottom surface, we find that the intrinsic vortex statistics is described by the $[U(1)_2 \times U(1)_2]\times U(1)_{-2}$ theory with anyons $I_{n_1, n_2, k}$, where $n_1 = 0, 1$, $n_2 = 0, 1$ are charges under $U(1)_2 \times U(1)_2$, and the vorticity $k$ coincides with the charge under $U(1)_{-2}$. $n_1, n_2, k$ satisfy the constraint $n_1 + n_2 + k = \rm{even}$. The topological spins of $I_{n_1, n_2, k}$  given by $\theta_{n_1, n_2, k} = \frac{\pi}{2} (n^2_1 + n^2_2 - k^2)$  are displayed in Table \ref{table_nu8_vortex}. The resulting vortex statistics is invariant under shifting $k \to k + 2$.

As in the $\nu = 4$ case, the above vortex statistics are exactly the same as in a 2D s-wave superconductor. Indeed, we can write the vortices as $X_k$, with $X$ running over the toric code anyons $\{I, e, m, \psi\}$, and $k$ denoting the vorticity. The topological spin $\theta_{X_k} = \theta_X$. Vortices with $k$-even have $X = \{1, \psi\}$ and vortices with $k$-odd have $X = \{e, m\}$. To match this description with the $[U(1)_2 \times U(1)_2] \times U(1)_{-2}$ theory, identify: $I_{0, 0, k} = I_k$, $I_{1,1, k} = \psi_k$, for even vorticity $k$, and $I_{1,0, k} = e_k$, $I_{0,1, k} = m_k$, for odd vorticity $k$.

\begin{table}[t]

\beq
\begin{array}{|c|c|c|}
\hline
n_1\,n_2\downarrow\, k\rightarrow     & 0 & 1  \\ \hline
0\,0\quad  (I)   & +1 &         \\ \hline
0\,1 \quad (e) &  & +1     \\ \hline
1\,0\quad (m) & & +1    \\ \hline 
1\,1 \quad (\psi) &  -1&\\ \hline
\end{array}
\eeq
\beq
\begin{array}{|c|c|c|c|c|}
\hline
\, k\rightarrow     & 0 & 1  \\ \hline
S^2  & +1 & -1       \\ \hline
\end{array}
\nn
\eeq

\caption{Vortices on the surface of the $\nu=8$ TSc. Top: Topological spins $e^{i \theta}$. The statistics are described by a $U(1)_{2}\times U(1)_{2} \times U(1)_{-2}$ theory or equivalently by a toric code theory. Here the column index is the flux $k\frac{hc}{2e}$, which coincides with the $U(1)_{-2}$ charge. The row index is the $U(1)_2 \times U(1)_2$ charge (or equivalently the toric code charge). $S$ time reversal maps all vortices into themselves. 
Bottom: $S^2$ assignments of vortices.}
\label{table_nu8_vortex}
\end{table}
We now specify the action of $S$ time reversal symmetry on the vortices. A vortex of strength $k$ now carries $4 k$ chiral Majorana modes (same as the edge of a $\nu = 4 k$ TSc in class BDI). Thus, the Bogolioubov quasiparticle $\psi_0$ is Kramers singlet ($S^2 = + 1$); $k = 1$ vortices transform into themselves under $S$ ($S:\, e_1 \to e_1,\,\, m_1 \to m_1$) and are (ordinary) Kramers doublets ($S^2 = -1$). Vortex transformation properties under $S$ are invariant under shifting $k \to k + 2$.

Note that since the Hall conductivity of the $T$-broken surface is integer-valued, magnetic monopoles in the bulk carry integer charge. Observe that vortices with flux $hc/e$, which correspond to bulk monopole insertions, are Kramers singlets ($S^2 = +1$). Hence, the neutral monopole must have $T^2 = +1$. Furthermore, a neutral monopole insertion must nucleate a vortex which has trivial mutual statistics with all other vortices. Thus, a neutral monopole nucleates the boson $I_2$. Therefore, the monopole is a boson. So we conclude that the monopole in this case is completely trivial.

Next, we consider condensing the strength 2 vortices, $I_{2}$, which are trivial both topologically and under $S$. This leads to charged boson excitations $e^{im\phi}$ with charge $m\frac{2e}{2}$. The charge $2e$ boson is identified with a Cooper pair, so only $m = 0, 1$ are topologically distinct. The quasiparticles now are $X_{k} e^{im\phi}$, $k = 0, 1$, $m = 0, 1$, where we use the toric code notation $X = \{I, e, m, \psi\}$. The statistics now includes a term that describes braiding of flux and charge, e.g. $\theta_{X_{k,\,m}}=\theta_{X}-\pi m k$. The electron is identified with the charge $e$ fermion $\psi_0 e^{i\phi}$, and has trivial mutual statistics with all other excitations. The resulting topological order has 8 particles and can be written as
\beq T^{\nu  = 8} = \{I_0,\, e_1,\, m_1,\, \psi_0\} \times \{I, f\} = \mathrm{Z}_2 \times \{I, f\}\eeq
Thus, $T^{\nu = 8}$ is a simple toric code topological order, as one would obtain from double vortex condensation in a 2D s-wave superconductor. The transition back to the superfluid phase occurs via condensation of $e^{i \phi}$. 

Time reversal acts on the anyons in the following way. The neutral fermion $\psi_0$ is a Kramers singlet ($T^2 = S^2 = +1$). The neutral vortex descendants $e_1$, $m_1$ are Kramers doublets ($T^2 = S^2 = -1$).  The charge $e$ boson $e^{i \phi}$ transforms as, $T: e^{i \phi} \to e^{-i \phi}$, and has $S^2 = + 1$ and $T^2 = -1$. Therefore, the electron $f = \psi_0 e^{i \phi}$ has $T^2 = -1$, as required. 

We see that the neutral anyons $\{I_0,\, e_1,\, m_1,\, \psi_0\}$ form precisely the eTmT toric code state in table \ref{tbl:eTmT}, so $T^{\nu = 8} = {\rm {eTmT}} \times \{I, f\}$. Note that the extra $U(1)$ symmetry plays little role in the STO: the electric charge is carried solely by the physical electron $f$ (and its fusion products with neutral anyons). Breaking this $U(1)$ symmetry amounts to ignoring the charge quantum number. Thus, the STO of the $\nu = 8$ TSc is the same as the STO of a 3D bosonic SPT with $T$-symmetry.

\section{Fermionic Kramers doublets}
\label{app:fKramers}
In this appendix, we prove the assertion in Eq.~(\ref{eq:fKramers}).  The proof is identical to one given by Ref.~\onlinecite{Levin2012} in the context of ordinary Kramers degeneracy. Starting with Eq.~(\ref{eq:T1T2ferm}), let us define $A = T^2_1$, $B = T^2_2$. Now, 
\bea B A |v\rangle &=& T c_1 T c_1 T c_2 T c_2 |v\rangle =    T c_1 T c_1 T^{\dagger} (-1)^F c_2 T c_2 |v\rangle = (-1)^{N_v} T c_1 (T c_1 T^{\dagger}) c_2 T c_2 |v\rangle \nn\\
&=& (-1)^{N_v + 1} T c_1 c_2 T c_1 c_2 |v\rangle \eea
where $(-1)^{N_v}$ is the fermion parity of $|v\rangle$. In the last step, we've used the fact that $T c_1 T^{\dagger}$ is a local fermion operator with support near point $1$, so $\{ T c_1 T^{\dagger}, c_2 \} = 0$. Now from Eq.~(\ref{eq:fDoublet}), $T c_1 c_2 |v\rangle = (-1)^{N_v} |v\rangle$. Therefore,
\beq B A |v\rangle = (-1)^{N_v + 1} |v\rangle \label{eq:BAv}\eeq
Now, 
\beq A = T c_2 T c_2 = - (-1)^F (T c_2 T^{\dagger}) c_2 = (-1)^F \tilde{A}, \quad B = T c_1 T c_1 = - (-1)^F (T c_1 T^{\dagger}) c_1 = (-1)^F \tilde{B} \label{eq:ABloc}\eeq
Here $\tilde{A} = - (T c_2 T^{\dagger}) c_2$ and $\tilde{B} = - (T c_1 T^{\dagger}) c_1$ are bosonic operators localized near points $2$ and $1$, respectively. Since $|v\rangle$ has short-range correlations, 
\beq \langle v| B^{\dagger} B|v\rangle \langle v | A^{\dagger} A |v\rangle  = \langle v|B^{\dagger} B A^{\dagger} A |v\rangle = \langle v| A^{\dagger} B^{\dagger} B A |v\rangle = || B A |v\rangle||^2 = 1 \label{eq:BBAA} \eeq
where we've used $[A, B] = [A^{\dagger}, B] = 0$ and Eq.~(\ref{eq:BAv}). Now,
\beq \langle v| A^{\dagger} A|v\rangle \ge |\langle v|A |v\rangle|^2, \quad   \langle v| B^{\dagger} B|v\rangle \ge |\langle v|B |v\rangle|^2 \label{eq:ineq}\eeq
with equality holding only if $A |v\rangle = \xi_1 |v\rangle$ , $B |v\rangle = \xi_2 |v\rangle$.  So, from Eq.~(\ref{eq:BBAA}),
\beq 1 =  \langle v| B^{\dagger} B|v\rangle \langle v | A^{\dagger} A |v\rangle \ge | \langle v|B|v\rangle \langle v|A|v\rangle|^2 = |\langle v| B A|v\rangle|^2 = 1\eeq
where in the second to last step we've used Eq.~(\ref{eq:ABloc}) and the fact that $|v\rangle$ has short-range correlations. Thus, the inequality (\ref{eq:ineq}) is saturated, so
\beq T^2_1 |v\rangle = A|v\rangle = \xi_1 |v\rangle, \quad T^2_2 |v\rangle = B|v\rangle = \xi_2 |v\rangle \label{eq:Eigen} \eeq 
By Eq.~(\ref{eq:BAv}), we have $\xi_1 \xi_2 = (-1)^{N_v + 1}$. Furthermore, 
\beq T_1 |v\rangle = T c_2 |v\rangle = T c_2 T^{\dagger} c_1 c_2 |v\rangle = -c_1 T c_2 T^{\dagger} c_2 |v\rangle = (-1)^{N_v} c_1 T^2_1 |v\rangle = 
(-1)^{N_v} \xi_1 c_1 |v\rangle \label{eq:T1v}\eeq
where we've used $T |v\rangle = c_1 c_2 |v\rangle$ in the first step. By taking the norm of the above equation and using $||c_1 |v\rangle||^2 = ||c_2 |v\rangle||^2 = 1$, we conclude $|\xi_1| = 1$. Now, from Eqs.~(\ref{eq:Eigen}), (\ref{eq:T1v}),
\beq T_1 T^2_1 |v\rangle = T_1 (\xi_1 |v\rangle) = \xi^*_1 T_1 |v\rangle = (-1)^{N_v} c_1 |v\rangle \eeq
On the other hand,
\beq T^2_1 T_1 |v\rangle = (-1)^{N_v} \xi_1 T^2_1 c_1 |v\rangle = - (-1)^{N_v} \xi_1 c_1 T^2_1 |v\rangle = -(-1)^{N_v} \xi^2_1 c_1 |v\rangle\eeq
where in the second to last step we've used Eq.~(\ref{eq:ABloc}), from which $\{T^2_1, c_1\} = 0$ follows. Thus, we conclude $\xi^2_1 = -1$, $\xi_1 = \pm i$. So, $\xi_2 = (-1)^{N_{v}+1} \xi^{-1}_1 = (-1)^{N_v} \xi_1 = \pm i$.

It remains to show that the four states $|v\rangle$, $T_1 |v\rangle= \xi_2 c_1 |v\rangle$, $T_2 |v\rangle = \xi_1 c_2 |v\rangle$, $T |v\rangle = c_1 c_2 |v\rangle$ are degenerate in energy and cannot be split by any local $T$-invariant perturbation $H$. Let's show this for $|v\rangle$ and $T_1 |v\rangle$ (the argument for the other states is similar). First, for any $H$, $\langle v|H| T_1 v\rangle = 0$, since $|v\rangle$ and $T_1 |v\rangle$ have different fermion parity. Next to show $\langle v|H|v\rangle = \langle T_1 v| H | T_1 v\rangle$, let us break up $H$ into a piece with support near point $1$, $H_1$, and a piece with support far away from point $1$, $H_{\rm{far}}$. Then,
\beq \langle T_1 v|H_1 |T_1 v \rangle = \langle v| c^{\dagger}_2 T^{\dagger} H_1 T c_2 |v\rangle = \langle v| c^{\dagger}_2 H_1 c_2 |v\rangle = 
\langle v|H_1 c^{\dagger}_2 c_2 |v\rangle = \langle v| H_1 |v\rangle \langle v| c^{\dagger}_2 c_2 |v\rangle= \langle v|H_1 |v\rangle \eeq
where we've used the $T$-invariance of $H_1$, the fact that $H_1$ is localized far from point $2$ and the fact that $|v\rangle$ has short-range correlations. As for $H_{\rm{far}}$,
\beq \langle T_1 v| H_{\rm{far}} |T_1 v\rangle = \langle v| c^{\dagger}_1 H_{\rm{far}} c_1 | v\rangle =\langle v|  H_{\rm{far}} c^{\dagger}_1 c_1 | v\rangle = \langle v|  H_{\rm{far}} |v\rangle \langle v| c^{\dagger}_1 c_1 |v \rangle = \langle v|  H_{\rm{far}} |v\rangle\eeq

\bibliography{ProjS}

\end{document}